\documentclass[12pt]{JHEP3}

\usepackage[OT2,OT1]{fontenc}
\newcommand\cyr{%
\renewcommand\rmdefault{wncyr}%
\renewcommand\sfdefault{wncyss}%
\renewcommand\encodingdefault{OT2}%
\normalfont
\selectfont}
\DeclareTextFontCommand{\textcyr}{\cyr}
\usepackage{amsmath}
\usepackage{amsfonts}
\usepackage{amssymb}
\usepackage{graphicx}
\usepackage{bm}

\def\be{\begin{equation}}
\def\ee{\end{equation}}
\def\ba{\begin{eqnarray}}
\def\ea{\end{eqnarray}}
\def\bs{\begin{subequations}}
\def\es{\end{subequations}}

\def\rme{e}
\def\rmd{d}
\def\rmi{i}
\def\p{\partial}

\def\cS{{\cal S}}
\def\cL{{\cal L}}
\def\cB{{\cal B}}
\def\cO{{\cal O}}
\def\cI{{\cal I}}
\def\tphi{\tilde\phi}
\def\B{\Box}
\def\a{\alpha}
\def\b{\beta}
\def\s{\sigma}
\def\vp{\varphi}
\def\O{\Omega}
\def\ga{\gamma}

\def\erf{{\rm erf}}
\newcommand{\Eq}[1]{(\ref{#1})}

\title{String theory as a diffusing system}

\author{Gianluca Calcagni\\
Institute for Gravitation and the Cosmos, Department of Physics,\\ The Pennsylvania State University,\\
104 Davey Lab, University Park, Pennsylvania 16802, U.S.A.\\
\\
Max Planck Institute for Gravitational Physics (Albert Einstein Institute)\\
Am M\"uhlenberg 1, D-14476 Golm, Germany\\
E-mail: \email{calcagni@aei.mpg.de}}

\author{Giuseppe Nardelli\\
Dipartimento di Matematica e Fisica, Universit\`a Cattolica,\\
via Musei 41, 25121 Brescia, Italia\\
\\
INFN Gruppo Collegato di Trento, Universit\`a di Trento,\\
38100 Povo (Trento), Italia\\ 
E-mail: \email{nardelli@dmf.unicatt.it}}

\abstract{Recent results on the effective non-local dynamics of the tachyon mode of open string field theory (OSFT) show that approximate solutions can be constructed which obey the diffusion equation. We argue that this structure is inherited from the full theory, where it admits a universal formulation. In fact, all known exact OSFT solutions are superpositions of diffusing surface states. In particular, the diffusion equation is a spacetime manifestation of OSFT gauge symmetries.}

\keywords{Tachyon Condensation, String Field Theory}

\preprint{JHEP02(2010)093 \hspace{2cm} arXiv:0910.2160}
\begin{document}

%%%%%%%%%%%%%%%%%%%%%%%%%%%%%%%%%%%%%%%%%%%%%%%%%%%%%%%%%%%%%%%%%%%%%%%%%%%%%%%%%%%%%%%%%%%%%%%%%%%%%%%%%%%%%%%%%%%%%%%%%%%%%%%%%%%%%%%%%%%%%%%%%%%%%%%%%%%%%%%%%%%%%%%%%%%%%%%%%%%%%%%%%%%%%%%%%%%%%%%%%%%%%%%%%%%%%%%%%%%%%%%%%%%%%%%%%%%%%%%%

\section{Motivation}

String field theory is a non-perturbative approach aiming to describe, at least in some of its parts, the microscopical geometrodynamics of Nature. Two of its most widely studied incarnations are open string field theory \cite{Wi86a,KS1,KS2} (OSFT; e.g., \cite{ohm01,sen04,FK}) and boundary string field theory \cite{wit1,wit2,sha1,sha2} (BSFT; e.g., \cite{sen04}), both of which are playgrounds whereon to develop new physical and mathematical ideas.

In particular, the OSFT effective action of the tachyon field, associated with the decay of unstable brane configurations, is manifestly non-local, inasmuch as it contains an infinite number of spacetime derivatives through operators of the form $e^\B$, where $\B=\p_\mu\p^\mu$ is the target (that is, spacetime) d'Alembertian. It entails a novel type of dynamics \cite{EW2,MZ}, often too complicated to be solved except with perturbative or numerical methods of limited range of validity. Stimulated by this problem, analytical non-perturbative methods have been found, and sometimes rediscovered from not-so-recent literature, which allow to handle pseudo-differential operators and find solutions of the non-local effective dynamics.

One of these methods \cite{roll,cuta2,cuta3,cuta4,cuta5} introduces an auxiliary coordinate $r$ along which the system is made to evolve according to the
\emph{diffusion equation}
\be\label{difeq}
(\B+\p_r)\phi=0\,,
\ee
where $\phi=\phi(r,x)$ is a scalar field (in the OSFT case, the tachyon). The infinite number of degrees of freedom corresponding to the initial conditions of the non-local `Cauchy problem' is encoded in the continuous variable $r$. The main consequences of eq.~\Eq{difeq} are that (i) dynamical equations become algebraic equations \cite{cuta2,cuta3,cuta5,cuta6}, (ii) the action of the $e^\B$ operator is a translation along the $r$ direction \cite{roll,cuta2,cuta3}, (iii) the spacetime dynamics is reduced to a well-defined local Cauchy problem \cite{cuta3}, and (iv) explicit solutions can be constructed. Regarding string theory, approximate analytic solutions were found for the Lorentzian bosonic and supersymmetric (susy) OSFT tachyon \cite{roll} and for the Euclidean supersymmetric OSFT tachyon \cite{cuta4,cuta5}, while an exact solution of the $p$-adic string was recovered \cite{cuta4}.

In parallel, some surprising relations were found between OSFT and BSFT solutions \cite{roll,cuta4}, which all pointed towards an interpretation of string theory as a diffusing system. Further support was gained in \cite{cuta5}, where the lower-level action of the OSFT supersymmetric tachyon was reconstructed starting from the diffusion equation for a scalar field with certain boundary conditions. 

All these results are based upon effective equations and, once these are given, the embedding within string theory can be forgotten. In doing so, however, it becomes increasingly difficult to explain and properly assess virtues and limitations of the method within the big picture of the full theory. The level of accuracy of solutions of OSFT equations of motion \cite{roll,cuta4,cuta5} and the brane tension ratio in a brane decay process \cite{cuta5} are so good that one wonders how this can happen considering that several approximations (level truncation, effective potential, and so on) are entailed.

It is the purpose of this paper to address these issues and discuss the link between the diffusion equation method and the universal structure of string field theory. Although approaches based on truncated actions have partly become obsolete because of the recent success in treating the full bosonic theory without integrating out the massive spectrum (see \cite{FK} for a review), the same techniques will make us better understand why and under what circumstances string field theory can be described, at the level of target embedding, as a diffusing system. Conversely, the heat-equation recipe for the construction of explicit tachyon solutions is potentially relevant also for different brane configurations or other fields of the string spectrum. %(lump solutions, for example, are manifestly background dependent). 
Moreover, it was instrumental for the construction of kink solutions interpolating different vacua of the theory \cite{cuta4,cuta5}. These solutions are still out of the scope of analytic techniques of modern OSFT, which have been applied to vacuum or marginally deformed configurations. Thereby, non-vacuum solutions with non-vanishing momentum are an open subject of study.

We will show that the diffusion equation \Eq{difeq} implements a gauge transformation at the level of the effective spacetime action, which allows one to construct non-trivial solutions from trivial configurations in a different frame. These configurations are the analogue of projector states in the conformal field theory (CFT) formulation, which are non-normalizable states formally satisfying the equation of motion. The target solutions found so far have a structure mimicking the exact analytic solutions of the non-truncated theory, corresponding to integral representations of the solution in terms of surface states. The relation between the CFT and spacetime results suggests novel applications of the same techniques. We give an example by arguing that also perturbative (marginal deformations) and non-perturbative (lumps, kinks) geometric configurations admit an integral representation in the bosonic and supersymmetric full theory.

The paper is organized as follows. In section \ref{rev} we review bosonic OSFT and its analytic solutions from the point of view of the string worldsheet. The spacetime effective theory and the solutions found with the diffusion equation method are discussed in section \ref{stet}, which also contains new material (section \ref{integr}) about the integral representation of solutions. The latter, eq.~\Eq{rhoin}, plays a crucial role in establishing the link between the full theory and the spacetime framework. In section \ref{sads} we discuss how spacetime effective solutions are expected to obey the diffusion equation on the grounds of the conformal properties of the exact solutions. The first goal is to argue from the properties of full OSFT that \emph{spacetime} solutions are expected to be diffusing, while so far the diffusion equation method has been just a useful trick without any such motivation. A second goal is to explain why certain initial conditions of the diffusion equation did not work as well as others. Concrete novel applications of these results to SFT and other non-local models are described in section \ref{appl}. The only solution constructed both in the target effective system and the full theory is the bosonic rolling tachyon with wild oscillations (marginal deformations). We draw a new detailed comparison between the diffusing and the exact solution in section \ref{wild}. Examples of finite superpositions of solutions are given in section \ref{fsup}, while sections \ref{isup} and \ref{pol} are devoted to solutions of non-local toy models. Future directions are outlined in section \ref{conc}.

%%%%%%%%%%%%%%%%%%%%%%%%%%%%%%%%%%%%%%%%%%%%%%%%%%%%%%%%%%%%%%%%%%%%%%%%%%%%%%%%%%%%%%%%%%%%%%%%%%%%%%%%%%%%%%%%%%%%%%%%%%%%%%%%%%%%%%%%%%%%%%%%%%%%%%%%%%%%%%%%%%%%%%%%%%%%%%%%%%%%%%%%%%%%%%%%

\section{OSFT}\label{rev}

%%%%%%%%%%%%%%%%%%%%%%%%%%%%%%%%%%%%%%%%%%%%%%%%%%%%%%%%%%%%%%%%%%%%%%%%%%%%%%%%%%%%%%%%%%%%%%%%

\subsection{OSFT action and tachyon}\label{osr}

In $\alpha'=1$ units, the OSFT bosonic action is \cite{Wi86a}
\be\label{SFT}
S=-\frac{1}{g_o^2}\int \left(\frac{1}{2} \Psi* Q\Psi+\frac13\Psi*\Psi*\Psi\right),
\ee
where $g_o$ is the open string coupling, * is a non-commutative product describing the gluing interaction of open strings, $Q$ is a BRST operator, and the string field $\Psi$ is a linear superposition of states whose coefficients correspond to the particle fields of the string spectrum. The open string field equation of motion is
\be\label{eom}
Q\Psi+\Psi*\Psi=0\,.
\ee

Contrary to the bosonic open string, there are several proposals for open superstring field theory, the first being Witten's \cite{wi86b,con1,con2,con3,con4,DR}. The action was later modified in \cite{PTY,AMZ1,AMZ2}: on a single non-BPS D$p$-brane, 
\ba
S&=&-\frac1{g_o^2}\int Y_{-2}\left(\frac1{2}\Psi_+* Q\Psi_++\frac13\Psi_+*\Psi_+*\Psi_++\frac1{2}\Psi_-* Q\Psi_--\Psi_+*\Psi_-*\Psi_-\right),\nonumber\\\label{sSFT}
\ea
%\ba
%S&=&-\frac1{g_o^2}\int Y_{-2}\left(\frac1{2}\Psi_+* Q\Psi_++\frac13\Psi_+*\Psi_+*\Psi_+\right.\nonumber\\
%&&\qquad\qquad\qquad\left.+\frac1{2}\Psi_-* Q\Psi_--\Psi_+*\Psi_-*\Psi_-\right),\nonumber\\\label{sSFT}
%\ea
where $Y_{-2}$ is a double-step inverse picture-changing operator and the string field $\Psi_\pm$ is a linear superposition of states (made of matter (super)fields $X^\mu$, $\psi^\mu$ and (super)ghosts $b$, $c$, $\b$, $\gamma$) in the GSO$(\pm)$ sectors \cite{Se99c}, respectively.

Although we will often recall results for the susy version of OSFT, for simplicity we concentrate on the bosonic case, eq.~\Eq{SFT}. This action is invariant under the infinitesimal gauge transformations
\be
\delta\Psi=Q\Psi+\Psi*\Lambda-\Lambda*\Psi\,,
\ee
where $\Lambda$ is a zero ghost number state. The gauge group of string theory is very large and, to the best of our knowledge, its full extent is still unknown. It includes spacetime diffeomorphisms and supersymmetry transformations \cite{QS}, as well as \emph{reparametrizations} of the open string (mappings of the string coordinate patch on the same region of the complex plane) \cite{ORZ}. In particular, when two states are related by a reparametrization, they are gauge equivalent. The converse may not be true, since one can define, e.g., one-parameter families of maps which are not reparametrizations for certain values of the parameter \cite{Ell2}.

In terms of the perturbative vacuum, the string field is a superposition of particle modes,
\be\label{pex}
\Psi \cong|\Psi\rangle=[\phi(x)+\dots]c_1|0\rangle\,,
\ee
where the first step indicates the state-vertex operator isomorphism, $c_1$ is a Laurent coefficient of the $c$ ghost, $c_1|0\rangle$ is the ghost vacuum with ghost number $-1/2$, and $x$ is the string center of mass. At the lowest truncation level, all particle fields in $\Psi$ are neglected except the tachyonic one $\phi$. The zero-momentum tachyon state $c_1|0\rangle$ belongs to the universal subalgebra ${\cal H}_{\rm univ}$ of the algebra of open string fields \cite{Sen99,RZ}.

As already pointed out from the very beginning \cite{Wi86a}, conservation of the BRST current implies invariance under reparametrizations of the open string, so one can make a partial gauge fixing and choose a particular parametrization, for instance one which locates the string midpoint at a convenient place in the conformal plane (by `convenient,' we mean one simplifying technical calculations).

For practical purposes one has to choose a CFT wherein to represent the string content. Different mappings of the string on this plane (\emph{conformal frames}) are possible \cite{FK,GRSZ2}, from upper half disks \cite{ohm01} to strips and cylinders \cite{FK}; these are associated with gluing procedures \cite{Sen99,RZ,LPP1,LPP2} which describe the Witten vertex. In particular, the representation of a string as a semi-infinite cylinder turned out to be very effective in the construction of an exact vacuum (translation invariant) solution in bosonic OSFT \cite{ORZ,Sch05,Oka06,FK1,RZ2,ES,TT} and in Berkovits' \cite{FK3} and cubic (polynomial) superstring field theories \cite{Er07b,AGM1,AGGKM,AGM2}. Different CFT's in the same conformal frame can describe different geometries; in this sense the CFT formulation of OSFT can be regarded as background independent \cite{Se90a,SZ93}.

%%%%%%%%%%%%%%%%%%%%%%%%%%%%%%%%%%%%%%%%%%%%%%%%%%%%%%%%%%%%%%%%%%%%%%%%%%%%%%%%%%%%%%%%%%%%%%%%

\subsection{Wedge states and projectors}

An essential tool for the construction of OSFT solutions are \emph{surface states} \cite{GRSZ2},
of which \emph{wedge states} and the \emph{sliver} projector are the most studied examples \cite{RZ,KP,RSZ1,RSZ2,GT1,RSZ3,Mat01,GT2,FO,HT,Muk01,Moe01,MT,AGM,MS,Sch02,Oka02,Ohm02,MaM,BMST1,BMST2,BMT} (for the supersymmetric case, see \cite{AGM,MS,Ohm02}). Recent discussions on the subject are \cite{Sch05,Sch02,FK2}.

Wedge states $|W_r\rangle$, often indicated as $|r\rangle=|W_{r-1}\rangle$, are a commutative subalgebra (with zero ghost number) of the string field star algebra \cite{RZ}:
\be\label{wsmol}
|W_r\rangle*|W_q\rangle=|W_{r+q}\rangle\,.
\ee
In the CFT presentation on the unit disk, $|W_r\rangle$ corresponds to a wedge of the disk with opening angle $\pi r$ at the origin, while in the cylinder presentation it is a semi-infinite strip of width $\pi(r+1)/2$ (see \cite{FK,RSZ3}). The state $|W_0\rangle$ corresponds to the identity state $|\cI\rangle$ \cite{RZ,Sch02,GJ1,GJ2,Klu02,TT2,Dru03}, $|W_1\rangle$ is the $SL(2,\mathbb{R})$ invariant vacuum $|0\rangle$, $|W_2\rangle=|0\rangle*|0\rangle$, while $|W_\infty\rangle$ is the sliver state. %The identity and the sliver are infinite reparametrizations of the vacuum state $|0\rangle$ \cite{Sch02}.

Wedge states are made of an infinite superposition of eigenstates $(L^+)^n|0\rangle$ of $\cL_0$:
\be\label{wed}
|W_r\rangle=\rme^{\frac{1-r}{2}L^+}|0\rangle\,,\qquad r\geq 0\,,
\ee
where $L^+=\cL_0+\cL_0^\dagger$ and $\cL_0$ is the zero mode of the stress-energy tensor in the sliver conformal frame, defined in terms of Virasoro generators \cite{Sch05,Sch02}:
\be
\cL_0\equiv L_0-2\sum_{k=1}^\infty\frac{(-1)^k}{4k^2-1}L_{2k}\,.\nonumber
\ee
The perturbative vacuum $c_1|0\rangle$ is an eigenstate of $\cL_0$. In general, eigenstates of $\cL_0$ are formed by arbitrary powers of $\cL_0^\dagger$, $\cB_0^\dagger$ and $c$ ghosts, where
\be
\cB_0\equiv b_0-2\sum_{k=1}^\infty\frac{(-1)^k}{4k^2-1}b_{2k}\,.\nonumber
\ee 

Surface states are convenient because they entail a change of representation of the string field Fock space, from one where the interaction term in eq.~\Eq{eom} is complicated and the BRST charge $Q$ is diagonal to one where the former term is converted to a linear form (see the discussion on squeezed states in \cite{KP}). In eq.~\Eq{wed}, $r$ can be thought of as defining the size of a `probe' making a scale-dependent `measurement' on the Fock vacuum. Solutions are non-local inasmuch as they are composed of probes at all scales $r\geq 0$ (see below). Equation \Eq{wed} defines a solution of
\be\label{bide}
\left(2\p_r+L^+\right)|W_r\rangle=0\,,
\ee
which is a universal diffusion equation. By universal we mean that involves only Virasoro (and, later, ghosts) operators \cite{Sen99}. The inner product of a wedge state with a primary field $\cO_h$ of weight $h$ is proportional to the vacuum expectation value of the field with rescaled argument:
\be
\langle 0|\cO_h(z)|W_r\rangle=\left(\frac2r\right)^h \langle 0|\cO_h\left(\frac{2z}r\right)|0\rangle\,.
\ee
The rescaling of the argument of $\cO_h$ is typical of diffusing fields.

Projectors \cite{RZ,GRSZ2,Sc02b,Yan04}, in particular \emph{special projectors} \cite{RZ2}, were first introduced in vacuum string field theory (VSFT) \cite{RSZ1,RSZ0,BMS1,BMS2,BMMS}; they are defined as surface states whose corresponding Riemann surfaces feature the string midpoint in their boundary. This implies that they are idempotent states for the star product,
\be\label{ppp}
|P_\infty\rangle * |P_\infty\rangle = |P_\infty\rangle\,.
\ee
They are important as they define conformal frames wherein OSFT can be solved. Equation \Eq{ppp}, in fact, is the equation of motion of the matter sector of VSFT.

The sliver projector $|W_\infty\rangle$ is a solution representing the bosonic D25-brane \cite{RSZ2} but it can also be constructed for boundary CFT's associated with other geometric configurations (for instance, the D-instanton sliver \cite{GT1,Moe01,MT}). It does not have finite norm but is instrumental for the construction of analytic solutions.

The sliver is only one of an infinite set of projector states, but others have been constructed, for instance the butterfly state \cite{RZ,GRSZ2,Sc02b,GRSZ1,Ok03a,Ok03b} and the nothing state \cite{RZ,GRSZ2,Sc02b}. In general \cite{RZ2}, one can define surface states which obey eq.~\Eq{wsmol},
\be\label{surfsd}
|P_r\rangle=\rme^{-\frac{r}{2}L^+}|{\cal I}\rangle\qquad r\geq 0\,,
\ee
where $L^+=(\cL_0+\cL_0^\star)/s$, $s>0$, and $[\cL_0,\cL_0^\star]=s(\cL_0+\cL_0^\star)$. $\star$ denotes BPZ conjugation ($L_n\to (-1)^nL_{-n}$), which coincides with Hermitian conjugation for all twist-even projectors (such as the sliver, $s=1$). The diffusion equation \Eq{bide} is unaltered, and $\cL_0|P_1\rangle=0$; in particular, $|P_1\rangle=|0\rangle$ in the sliver-based family of wedge states. $|P_r\rangle$ are wedge states if $L^+$ is defined in the conformal frame of the sliver. There (cylinder presentation), the operator $\rme^{-qL^+}$ creates a semi-infinite strip of width $\pi q$. Pure-gauge solutions do not contain a component along the vacuum state $|P_1\rangle$ (in this sense, their normalization is arbitrary). The only common state of different families of surface states is the identity $|P_0\rangle=|\cI\rangle$. Since $\cL_0|P_\infty\rangle=0$ \cite{RZ2}, $\cL_0$ is the zero-mode Virasoro operator in the special projector conformal frame. In these conformal frames the string midpoint is on the boundary and the left and right half-strings behave as independent objects. 

Different projector frames associated with the same boundary CFT give gauge-equivalent solutions. To any twist-invariant, single-split projector there corresponds a solution in a different gauge; special projectors yield simpler solutions \cite{RZ2}. Projectors are related with each other (in particular, the sliver) by finite midpoint-preserving reparametrizations (i.e., large gauge symmetries) of the open string coordinate $\tau$. Corresponding to a reparametrization $\tau'=\vp(\tau)$, there is an operator $U_\vp$ acting on the space of string fields and representing an OSFT gauge transformation, $|W_r\rangle= U_\vp |P_r\rangle$.

Within a given family of surface states, there is also a reparametrization leaving the projector invariant and mapping the states into one another, $|P_r\rangle\to |P_{\rme^\beta r}\rangle$, where $\beta$ is real. This corresponds to a conformal rescaling ${\cal P}_r\to {\cal P}_{r'}$ of the associated one-punctured disk in the presentation where the local coordinate patch is that of the projector $|P_\infty\rangle$. In the sliver family, any wedge state can be made to approach the sliver \cite{ORZ,GRSZ1}, although regular surface states cannot be mapped to projectors by a finite reparametrization as they define different topologies.

The main features of the solutions of OSFT are shown by a toy model with zero ghost number. The general solution of its equation of motion
\be\label{pippo}
(\cL_0-1)\Psi+\Psi*\Psi=0
\ee
is \cite{RZ2}
\be\label{soluP}
|\Psi_s\rangle=|P_\infty\rangle+\int_0^\infty \rmd r \mu_s(r) L^+ |P_r\rangle\,,
\ee
where $\mu_s$ is a one-parameter family of measures. Projectors automatically solve eq.~\Eq{pippo}. %Solutions with different values of $s$ are not related by a midpoint-preserving reparametrization. 
The number of special projectors for a given $s$ is argued to be finite. Examples are the sliver ($s=1$), the butterfly and the moth ($s=2$). In the first case, $2\mu_1=\sum_{n=1}^\infty \delta(r-n)$. Otherwise, the integral measure is 0 for $r<1$.
The solution can be also recast as
\be
|\Psi_s\rangle= f_s(L^+) |\cI\rangle\,,\nonumber
\ee
where $f_s(x)=[{}_1F_1(1,1+1/s,x/2)]^{-1}$. The measure $\mu_s(r)$ is the inverse Laplace transform of $f_s(2x)/(2x)$. It is amusing that the Kummer function ${}_1F_1$ appears both in the full zero-ghost-number theory (as a functional of the Virasoro operator $L^+$) and, in special cases, also in the effective target system \cite{cuta2,cuta4} (as lowest-level solutions of the effective equation with non-local operators).

As a superposition of surface states, the solution eq.~\Eq{soluP} `interpolates' between the identity state and the special projector. We shall see that this is the same structure of target effective solutions.

%%%%%%%%%%%%%%%%%%%%%%%%%%%%%%%%%%%%%%%%%%%%%%%%%%%%%%%%%%%%%%%%%%%%%%%%%%%%%%%%%%%%%%%%%%%%%%%%

\subsection{Vacuum revolution}\label{vacr}

In order to construct solutions to the OSFT equation of motion, the simple subalgebra of wedge states must be modified introducing ghost operators. The zero-momentum tachyon vacuum solution was first found by Schnabl \cite{Sch05} as a sum over wedge states with insertions, $\psi_r$:
\be\label{vacs}
\Psi=\psi_\infty-\sum_{r=0}^\infty\p_r\psi_r\,.
\ee
This solution is normalizable, as the divergence from the sliver cancels out the one from the infinite sum of states. Notice that wedge states with insertions can be written as \cite{Oka06} $\psi_r=2 c_1|0\rangle*|W_{r-1}\rangle*B^+c_1|0\rangle$, where $B^+={\cal B}_0+{\cal B}_0^\dagger$, so they obey the diffusion-type equation
\be\label{difeqgh}
\p_r\psi_r+c_1|0\rangle*|W_{r-1}\rangle*B^+L^+c_1|0\rangle=0\,,\qquad r\geq 1.
\ee

As it is constructed on a specific conformal frame (the sliver's), this solution is frame dependent but this does not result in any loss of generality, as Schnabl's solution can be built on other projector frames \cite{ORZ}. Like eq.~\Eq{bide}, it is universal but background dependent, in the sense that it is formulated only in terms of Virasoro and ghost operators which describe a particular brane configuration/CFT.

Schnabl's solution obeys the gauge condition ${\cal B}_0 \Psi=0$, which does not fix the gauge completely \cite{Ell1}. Indeed, the second piece of the solution can be written as the limit of a pure gauge state satisfying the ${\cal B}_0$ gauge \cite{Oka06} (star product is understood):
\be\label{sep}
\sum_{r=0}^\infty \p_r\psi_r=\lim_{\lambda\to 1}\Psi_\lambda=\lim_{\lambda\to 1}\lambda (Q\Phi)\frac{1}{1-\lambda\Phi}=\lim_{\lambda\to 1}\rme^{-\Lambda(\lambda)}Q\rme^{\Lambda(\lambda)}\,,
\ee
where $\Lambda(\lambda)=-\ln(1-\lambda\Phi)=\sum_{n=1}^\infty (\lambda\Phi)^n/n$ and $\Phi=B^+c_1|0\rangle$ is the tachyon vacuum with ghost operator $B^+$. When $\lambda<1$, eq.~\Eq{sep} is gauge equivalent to $\Psi_\lambda=0$. The construction of non-trivial solutions from pure gauge configurations was formalized in \cite{Oka06,ES,Er07a,FKP} (see also \cite{AGGKM,AGM2} for the vacuum solution of cubic open superstring field theory).

%%%%%%%%%%%%%%%%%%%%%%%%%%%%%%%%%%%%%%%%%%%%%%%%%%%%%%%%%%%%%%%%%%%%%%%%%%%%%%%%%%%%%%%%%%%%%%%%

\subsection{Marginal deformations}

When the worldsheet action of the CFT is deformed by an exactly marginal operator $J$ at the boundary, one obtains a one-parameter family of boundary conditions which represents a dynamical perturbation of the same geometric configuration, typically a D$p$-brane \cite{Se90a,ReS,SZ00,IN,TaT,Klu03,KiT}. Wilson lines and the rolling tachyon are examples of marginal deformations. Perturbative tachyon solutions for exactly marginal deformations were constructed for bosonic OSFT \cite{FK3,Er07a,FKP,Sch07,KORZ,KO1,LPT,Kwo08,Kis08} and Berkovits' OSFT \cite{FK3,Er07a,Ok07a,Ok07b,KO2}. These rolling tachyon solutions are interpreted as the beginning (or the end, if the tachyon vacuum is perturbed \cite{Kwo08}) of a brane decay, but they do not capture the whole dynamical process since the initial and final geometric configurations of the decay are very different in terms of the underlying boundary CFT. In particular, lumps and kinks are not exactly marginal deformations, and we do not expect them to be described by a perturbative series in the deformation parameter $\lambda$.

Let $J(z)=c\cO(z)$, where $\cO$ is a dimension-one matter primary operator. The operator $\cO(z)=\rme^{X^0(z)}$ describes a tachyon field which starts rolling from the unstable vacuum at $x^0=-\infty$ towards the non-perturbative vacuum \cite{Se02a,Se02b}. The time-dependent bosonic solution is \cite{Sch07,KORZ}
\be\label{mads}
\Psi_\lambda=\sum_{n=1}^{+\infty} \lambda^n \psi_n\,,
\ee
where $\psi_1=J(0)|0\rangle$ is a solution of the linear equation $Q\psi_1=0$ and $\psi_n$ are wedge states with $n$ insertions of $J$ on their boundary in the sliver frame, and are determined recursively from $\psi_1$. This is almost of the same form as the vacuum solution \Eq{vacs}, with the difference that the latter is pure gauge when $\lambda<1$, ill-defined when $\lambda>1$, and physical when $\lambda=1$, while eq.~\Eq{mads} is a family of physical solutions also for $\lambda<1$. This family of solutions respects the Schnabl gauge if $J$ has regular operator product expansion (OPE), otherwise one must add perturbative counterterms which violate it. In the superstring case, $\cO$ is a superconformal primary field of dimension $1/2$ \cite{Er07a,Ok07a} and, due to the form of Berkovits' equation of motion, the solutions also happen to be pure gauge configurations from the perspective of bosonic OSFT.

In \cite{FKP} an interesting variant of the above construction was proposed, where one starts with an exact solution in a `large' Hilbert space (where states are in general non-normalizable) and defines a singular gauge transformation which pushes this solution into a perturbative solution in the physical Hilbert space. In other words \cite{Kwo08}, the solution is of the form $\Psi_\lambda= \rme^{-\Lambda(\lambda)}Q\rme^{\Lambda(\lambda)}$. When $\rme^{\Lambda}\approx {\cal I}+\Lambda$ can be deformed to the identity state, $\Psi_\lambda$ is a pure gauge solution, otherwise $\Lambda$ defines a large (even singular) gauge transformation and the solution becomes physical. The procedure of \cite{FKP} makes use of integrated vertex operators and a general construction was obtained in \cite{KO1}. The solutions of \cite{Sch07,KORZ} (non-integrated vertex operators) and \cite{FKP,KO1} are all gauge equivalent \cite{Kis08}.

Diffusing states will be important in what follows and it is worth noticing that solutions with marginal deformations are constructed in terms of generalized wedge states \cite{KO1}. These are defined as
\be
|U_r\rangle=\sum_{n=0}^\infty \lambda^n |U_r^{(n)}\rangle=|W_r\rangle+O(\lambda)\,,
\ee
and they obey the usual composition rule $U_r \bullet U_q\equiv U_r * U_0^{-1}* U_q= U_{r+q}$ under a deformed star product $\bullet$. These states are closed for the BRST operator ${\cal Q}$ on the deformed background, ${\cal Q} |U_r\rangle=0$.

%%%%%%%%%%%%%%%%%%%%%%%%%%%%%%%%%%%%%%%%%%%%%%%%%%%%%%%%%%%%%%%%%%%%%%%%%%%%%%%%%%%%%%%%%%%%%%%%

\subsection{String field theories and gauge equivalence}\label{gau}

We conclude this section by reviewing the relations between different solutions in the same SFT and solutions of different SFT's. Regarding the former, it was established in \cite{Ell2} that, if the BRST operator ${\cal Q}$ around the tachyon vacuum has no cohomology at any ghost number, every solution $\Psi$ of bosonic OSFT can be written as a formal gauge transformation of the tachyon vacuum,
\be\label{psu}
\Psi = U^{-1}{\cal Q}U\,.
\ee
If the transformation $U$ is singular (or, perhaps, just large), the solution is physical. This happens when $U$ annihilates a rank-one projector $|P_\infty\rangle$ of the star algebra, i.e., when $|P_\infty\rangle$ is in the right kernel of $U$, $U|P_\infty\rangle = 0$. The examples considered in \cite{Ell2} include all known cases: the non-perturbative vacuum ($|P_\infty\rangle=|0\rangle$, by definition), the perturbative vacuum ($|P_\infty\rangle=|W_\infty\rangle$, sliver), marginal deformations with trivial OPE \cite{Sch07,KORZ} ($|P_\infty\rangle=|U_\infty\rangle$, generalized sliver of \cite{KO1}) and marginal deformations with non-trivial OPE \cite{FK3,FKP,KO1} ($|P_\infty\rangle=|U^{-1/2}U_\infty U^{-1/2}\rangle$).

As far as different theories of interacting strings are concerned, solutions of OSFT and BSFT can be mapped onto each other, as pointed out in \cite{roll,Ell2,Ell1,CST}. The marginal rolling tachyon solution can be mapped to the bounded BSFT solution, so that its wild oscillations \cite{MZ,roll,CST,FGN2} are interpreted as an artifact of a complicated time-dependent gauge transformation \cite{Ell1}. Within OSFT's, there also exists a mapping between supersymmetric and bosonic classical solutions \cite{FK3,Er07a}, cubic and Berkovits' supersymmetric SFT's \cite{AGM2,FK4}, and between different polynomial supersymmetric SFT's \cite{Kro09}. For these reasons, it is not restrictive to consider solutions of one particular SFT (for instance, those of eq.~\Eq{sSFT}) and, on the other hand, the physical interpretation of such solutions can be made more transparent when considering their duals in another theory.

By merging the above results, it is reasonable to expect that (a) the findings of \cite{Ell2} can be extended to superstring field theory and (b) solutions with and without marginal deformations can be treated on the same ground, eq.~\Eq{psu}. These are important points as the susy effective non-local system stemming from eq.~\Eq{sSFT} admits solutions which do not correspond to marginally deformed CFT's but have been constructed with the same method (diffusion equation) as for bosonic and susy solutions for marginal deformations. Therefore, it is natural to ask whether this method is actually a spacetime formulation or approximation of the gauge properties of the CFT framework. Once this question was answered, one could have a clearer guidance for future applications of the same techniques. We will argue the answer to be affirmative.

%%%%%%%%%%%%%%%%%%%%%%%%%%%%%%%%%%%%%%%%%%%%%%%%%%%%%%%%%%%%%%%%%%%%%%%%%%%%%%%%%%%%%%%%%%%%%%%%%%%%%%%%%%%%%%%%%%%%%%%%%%%%%%%%%%%%%%%%%%%%%%%%%%%%%%%%%%%%%%%%%%%%%%%%%%%%%%%%%%%%%%%%%%%%%%%%

\section{Effective spacetime action and diffusing solutions}\label{stet}

%%%%%%%%%%%%%%%%%%%%%%%%%%%%%%%%%%%%%%%%%%%%%%%%%%%%%%%%%%%%%%%%%%%%%%%%%%%%%%%%%%%%%%%%%%%%%%%%

\subsection{Lowest-level non-local spacetime actions}

Let us consider the bosonic action eq.~\Eq{SFT} introduced in section \ref{osr}. In Siegel gauge ($b_0\Psi=0$) and around the perturbative vacuum, the BRST operator is $Q= c_0L_0=(p^2-1)c_0+\dots$, where $L_0$ is the Virasoro zero mode of the total stress-energy tensor in the upper-half disk presentation and $-p^2$ is the Fourier transform of the d'Alembertian. At lowest truncation level in eq.~\Eq{pex}, one can write down an effective spacetime action for the tachyon, which exhibits a non-local interaction \cite{KS1,KS2}:
\be
\cS =\frac{1}{g_o^2}\int \rmd^D x \left[\frac{1}{2}\phi(\B+1)\phi
-\frac{\rme^{3r_*}}{3}\tphi^3\right],\label{bosa}
\ee
where $\tphi\equiv \rme^{r_*\B}\phi$ and
\be
r_*=\ln (3^{3/2}/4)\approx 0.2616\,.\label{rst}
\ee
The value of $r_*$ is dictated by conformal invariance (which partly survives although level truncation breaks gauge invariance even without explicit gauge fixing) and does not depend on the type of presentation chosen for the string worldsheet. It also appears in susy OSFT, for instance in the formulation given by eq.~\Eq{sSFT}.

There, the operator $Y_{-2}$ can be either chiral and local \cite{AMZ1,AMZ2} or non-chiral and bilocal \cite{PTY} (see the literature and the review \cite{ohm01} for details). These two theories predict the same tree-level on-shell amplitudes but different off-shell sectors.
The non-local effective action for the tachyon has been constructed for the non-chiral version \cite{PTY,AKBM,AJK}. In the 0 picture and at level $(1/2,1)$, which is the lowest for the susy tachyon effective action, all particle fields in $\Psi_\pm$ are neglected except the tachyonic one and an auxiliary level $-1$ field $u(x)$. This field is responsible for the emergence of a quartic effective potential for the tachyon. In fact, the Fock-space expansion of the string field is truncated so that the spacetime action on a D$p$-brane reads \cite{AKBM,AJK}
%\ba
%S&=&\frac1{g_o^2}\int \rmd^{p+1}x\left[\frac12 \phi\B\phi+\frac14\phi^2+u^2\right.\nonumber\\
%&&\left.-\frac{\rme^{2 r_*}}3(\rme^{r_*\B}u)(\rme^{r_*\B}\phi)^2\right]\,.\label{su}
%\ea
\be
\cS=\frac1{g_o^2}\int \rmd^{p+1}x\left[\frac12 \phi\B\phi+\frac14\phi^2+u^2-\frac{\rme^{2 r_*}}3(\rme^{r_*\B}u)(\rme^{r_*\B}\phi)^2\right]\,.\label{su}
\ee
Combining the equations of motion for $u$ and $\phi$, one obtains an equation for $\tphi$ alone:
\be\label{susa}
\left(\B+\frac12\right)\rme^{-2r_*\B}\tphi- \frac{\rme^{4r_*}}{9}\tphi\,\rme^{2r_*\B}\tphi^2=0\,.
\ee
The effective potential is extremely complicated and until recently \cite{cuta5} it has made it prohibitive to find even numerical solutions of this system.

%%%%%%%%%%%%%%%%%%%%%%%%%%%%%%%%%%%%%%%%%%%%%%%%%%%%%%%%%%%%%%%%%%%%%%%%%%%%%%%%%%%%%%%%%%%%%%%%

\subsection{Diffusion equation method}\label{diem}

The form of the effective equations \Eq{bosa} and \Eq{susa} triggered a considerable amount of work on non-local theories \cite{EW2,MZ,LV,GKL,beri0,CHY,VV,GKR,pro06,vla07,BK1,BK2}. The difficulties one meets when dealing with non-local operators are both interpretative and technical. On one hand, the construction of non-perturbative solutions is a highly non-trivial task also for a Minkowski metric. A truncation of $\rme^\B$ operators leads to a higher-derivative effective theory, which is arguably different from the original in all respects, unless certain unverifiable conditions (for instance, slow variation of the fields or convergence of perturbative series) are satisfied. On the other hand, the Cauchy problem is unclear, as one should specify an infinite number of initial conditions which would correspond to the knowledge of the solution (if analytic) around the initial point.

Among the attempts to address these issues, the diffusion equation method turned out to be a convenient mathematical tool. Its early applications to OSFT and the $p$-adic string
\cite{vol03,FGN,vla05} were soon followed by an extensive study of the dynamics of these diffusing systems with analytic, semi-analytic, or numerical techniques \cite{roll,cuta2,cuta3,cuta4,cuta5,cuta6,Jou07,Jo081,MuN,Jo082,NuM}. The method can be summarized as follows \cite{roll,cuta3,cuta4}.
\begin{itemize}
\item Interpret $r_*$ as a fixed value of an auxiliary evolution variable $r$, so that the scalar field $\phi=\phi(r,x)$ is thought to live in $1+D$ dimensions and evolve via the diffusion equation \Eq{difeq}.
\item Given the initial condition $\phi(0,x)$ at $r=0$, the solution of the diffusion equation is
\be\label{solde}
\phi(r,x)=\rme^{r\B}\phi(0,x)\,.
\ee
\item In particular, the effect of the non-local operator $\rme^{q\B}$ is a shift of the auxiliary variable $r$:
\be\label{tra}
\rme^{q\B}\phi(r,x)= \rme^{-q\,\p_r} \phi(r,x)=\phi(r-q,x)\,.
\ee
\item As a consequence, the system becomes \emph{local} in spacetime variables. For this reason the Cauchy problem is well-defined and one has to specify only a finite number of initial conditions \cite{cuta2,cuta3,BK1}. Intuitively, the infinite number of degrees of freedom of the non-local system have been transferred into a \emph{field} configuration at $r=0$. The $(1+D)$-dimensional system solved by some $\phi(r,x)$ is referred to as \emph{localized}.
\item The Hamiltonian and conjugate momenta are easily constructed from non-local Lagrangian systems. Quantization of the degrees of freedom, if desired, stems from a finite symplectic structure.
\item Not only the calculation of the energy-momentum tensor $T_{\mu\nu}$ for the localized system is much simpler than in the non-local case \cite{cuta1}, but it shows that the form of $T_{\mu\nu}$ is precisely the one for a diffusing scalar \cite{cuta3}. This is a self-consistent check that all known solutions of the original non-local model obey the diffusion equation.
\end{itemize}

The construction of explicit solutions goes through the following steps.
\begin{enumerate}
\item[(A)] Find a solution $\phi(0,x)$ of the corresponding \emph{local} system ($r=r_*=0$ everywhere). This is the initial condition for a system that evolves in $r$. 
\item[(B1)] If the initial condition is chosen to be constant almost everywhere,
the final configuration $\phi(r_*,x)$ obtained by diffusion along $r$ is a smooth function which solves (exactly or approximately) the original non-local system.
\item[(B2)] If the initial condition is chosen to be continuous, then:
	\begin{enumerate}
	\item[(B2a)] Solve the eigenvalue equation of the d'Alembertian operator, $\B G_p(x)=-p^2 G_p(x)$. The eigenfunctions $G_p$ are just plane waves in the Minkowski case, so the checklist below is easy to carry out for string theory \cite{roll}. It is much less trivial on curved backgrounds \cite{cuta2}. 
	\item[(B2b)] Write the local solution $\phi(0,x)$ as a linear combination (sum or integral) of the eigenfunctions of the d'Alembertian operator, e.g., $\phi(0,x)=\sum_p c_p G_p(x)$.
	\item[(B2c)] Look for non-local solutions $\phi(r,x)$ of the type $\phi(r,x)=\rme^{r\B} \phi(0,x)=\sum_p \rme^{-r p^2} c_pG_p(x)$, for some constant $r$.
	\end{enumerate}
\item[(C)] The constant $r_*$ and the normalization such that $\phi(r_*,x)$ is a solution (exact or approximate) can be found by looking at the asymptotic behaviours of the equation of motion.	
\end{enumerate}
Examples of a constant initial condition in the sense of distributions are the $p$-adic string \cite{cuta4}, supersymmetric OSFT \cite{cuta4,cuta5}, and some cosmological toy models \cite{cuta6}. In the first case, the equation of motion admits $\phi=0$ as constant solution, so that $\phi(0,x)=\delta(x)$ is a local solution everywhere except at the origin. The diffusion equation \Eq{difeq} smoothens it to a Gaussian lump; this solution is exact. In the second case, the initial condition is a step function corresponding to the position of the two local minima of the tachyon potential ($\phi=\pm 1$). Upon diffusion, this configuration evolves to a one-dimensional kink:
\be\label{erf}
\phi=\erf\left(\frac{x}{\sqrt{4r}}\right)
\ee
for some $r$, where erf is the error function. This solution is approximate with very good accuracy \cite{cuta5}. No such solution has yet been found in the full theory.

Examples of continuous initial conditions are the rolling tachyon with marginal deformations \cite{roll,FGN} and some cosmological toy models \cite{cuta2}. In the first case the bosonic tachyon profile is
\be\label{solmd}
\phi=\rme^{t}-\sum_{n=2}^{+\infty} (-1)^n c_n \rme^{n t}\,,
\ee
where $t=x^0$ is the time coordinate and \cite{roll,FGN2,FGN}
\be\label{psix}
c_n=6^{1-n}n \rme^{-(4n^2-9n+5)r_*}\,,
\ee
%\be
%c_n\sim 1.622\cdot 10^{-0.454 n(n-0.538)}\,,
%\ee
while for the non-chiral bilocal susy OSFT \cite{roll}
\be
\phi=3\sum_{n=0}^{+\infty}(-1)^n \rme^{-r_*(2n+1)^2} \rme^{(2n+1)t/\sqrt{2}}.\label{nloc3x}
\ee
Both solutions are approximate and display the well-known wild oscillations.

The pseudo-differential operator $\rme^\B$ may be seen also as a rescaling, rather than a translation, in $r$. This happens when the diffusion equation is of the form
\be\label{difeq2}
(\B+r\p_r)\phi=0\,,
\ee
where now the $r$ gradient is logarithmic. Then, defining $\varrho\equiv \ln r$,
\ba
\rme^{q\B}\phi(r,x)&=&\rme^{-q\p_\varrho}\phi(\rme^\varrho,x)\cr
&=&\phi(\rme^{\varrho-q},x)\cr
&=&\phi(\rme^{-q}r,x)\,.\label{resc}
\ea
We shall use this property to prove one the main results of the next section.

%%%%%%%%%%%%%%%%%%%%%%%%%%%%%%%%%%%%%%%%%%%%%%%%%%%%%%%%%%%%%%%%%%%%%%%%%%%%%%%%%%%%%%%%%%%%%%%%

\subsection{Solutions in integral form}\label{integr} 

At this point we stress a property of the solutions of the diffusion equation which have been often noticed informally. For simplicity we only discuss the one-dimensional homogeneous case, where the heat equation is (we adopt signature ${-}{+}{\cdots}{+}$)
\be\label{difeqt}
(\p_t^2-\a\p_r)\phi=0\,.
\ee
If $\a=+1$ (eq.~\Eq{difeq}), diffusion occurs towards the natural direction and the solution of \Eq{difeqt} is $C^\infty$. On the other hand, when $\a=-1$ one expects to find a singularity somewhere during the field evolution (see \cite{roll} below eq.~(35)). On general grounds, if a solution $\phi(r,t)$ of eq.~\Eq{difeqt} is not $C^\infty$, then its Wick-rotated version $\phi(r,\rmi t)$ (i.e., the Euclidean solution if the starting point was Minkowski) is a regular solution. Of course, it is not guaranteed that $\phi(r,\rmi t)$ is also a solution of the Wick-rotated equation of motion; in fact, this hope has been betrayed in all known cases \cite{roll,cuta4}.

Another way to recast these considerations is to construct a solution of eq.~\Eq{difeqt} in integral form, once the initial condition in $r$ is known. If the diffusion coefficient is positive, the standard procedure is based on the heat kernel
\be
\label{ker}
K(r,\s)=\frac{\rme^{-\frac{\s^2}{4r}}}{2\sqrt{\pi r}}\,.
\ee
The normalization is chosen so that 
\be
\label{no}
\lim_{r\to 0} K(r,\s)=\delta(\s)\,,
\ee
in the sense of distributions. Since $K(r,\s)$ is the solution of the heat equation
\be\label{ht}
(\p^2_\s-\p_r)K=0
\ee
with eq.~\Eq{no} as initial condition, any solution $\phi(r,t)$ of the heat equation with initial condition $\phi(0,t)$ can be easily obtained as the convolution of the heat kernel with the initial condition:
\be
\label{hkp}
\phi(r,t)=\int_{-\infty}^{+\infty} \rmd t'\, K(r,t-t')\, \phi(0,t')\,.
\ee
The simplest non-trivial example of (smooth) solution of this form is the error function eq.~\Eq{erf} ($x=t$), corresponding to the initial condition $\phi(0,t)={\rm sgn}(t)$.

If the diffusion coefficient is negative, the convolution with the heat kernel cannot be done directly in general. However, one can apply a different method. Let us consider an harmonic function $u$ in the variables ($t,\s$),
\be\label{har}
\nabla^2 u (t,\s)=\p^2_\s u + \p^2_t  u=0\,,
\ee
and let $\phi(0,t)=u(t,0)$. Then,
\be
\label{so}
\phi(r,t)=\int_{-\infty}^{+\infty} \rmd\s \, K(r,\s)\,u(t,\s)
\ee
is the solution of the diffusion equation \Eq{difeqt} with $\a=-1$ and initial condition $\phi(0,t)$, provided $u(t,\s)$ satisfies certain conditions. The initial condition is trivially satisfied by virtue of eq.~\Eq{no}. Then, $(\p^2_t+\p_r)\phi(r,t)=0$, where we have used eqs.~\Eq{har} and \Eq{ht} and integrated by parts twice. 

Therefore, the conditions required on $u$ are those that legitimate the double integration by parts, i.e., $u$ and its first $\sigma$ derivative do not possess singularities along the real $\sigma$ axis and, asymptotically, $u$ be polynomially bounded (then, $Ku$ tends to 
zero at $\s \to \infty$). The spiky solutions of \cite{roll} are indeed of this form; see eqs.~(23) and (40) therein, but with the limit taken in the strong sense. In the bosonic case, $u(t,\s)\propto (1+\cosh t \cos\s)/(\cosh t+\cos\s)^2$. At $t=0$ the denominator of $u$ develops poles in the integrand, which are responsible for the spike at the origin.
Also the wildly oscillating solutions (bosonic and susy) can be written in integral form, eq.~(45) of \cite{roll} (they are actually the analytic continuation of the spiky solutions with $t<0$). On the other hand, there is no regular harmonic function associated with either the error function or any other solution of the diffusion equation with distribution-like initial condition.\footnote{The Gaussian lump of the $p$-adic string is the other notable example \cite{cuta4}} After regularizing the initial condition, however, one can apply eq.~\Eq{har}, and eq.~\Eq{so} follows when removing the regulator.

Equations \Eq{hkp} and \Eq{so} are two different ways to write the solution of the diffusion equation. However, while eq.~\Eq{so} cannot be recast as eq.~\Eq{hkp}, the converse is true. Suppose to take eq.~\Eq{so} with $u$ being a function obeying the wave (rather than Laplace) equation. Then, the resulting $\phi$ is a solution of the diffusing equation with $\a=+1$, hence always regular. To pass from one case to the other, it is sufficient to make a Wick rotation $\s\to \rmi \s$ or $t\to \rmi t$.

To summarize, solutions of the diffusion equation \Eq{difeqt} always admit the integral representation eq.~\Eq{so}, where $u$ is harmonic in its variables and $u(t,0)=\phi(0,t)$. If $\a=-1$, $u=u(t,\s)$, while if $\a=+1$ one has $u=u(t,\rmi \s)$.

We now further manipulate eq.~\Eq{so} into a very useful form to be invoked later. Since $K$ is even in $\s$, one has
\be
\phi(r,t)= \int_{0}^{+\infty} \rmd\s\,\frac{\rme^{-\frac{\s^2}{4r}}}{\sqrt{\pi r}}\ w(t,\s)\,,\nonumber
\ee
where $w$ is (proportional to) the even part of $u$: 
\be\label{pip2}
2\,w(t,\s)= u(t,\s)+u(t,-\s)\,,\qquad w(t,0)=\phi(0,t).
\ee
Notice that
\begin{itemize}
\item[(a)] If $u$ obeys the Laplace or wave equation, so will $w$. In the first case,
\be\label{eqa}
\p_t^2w+\p_\s^2w=0\,.
\ee
\item[(b)] Since $\p_\s w$ is odd by construction, it vanishes at the origin:
\be\label{eqb}
\p_\s w(t,\s)\Big|_{\s=0}=0\,.
\ee
\end{itemize}
After the change of variable
\be
\rho\equiv \frac{\s^2}{4 r}\,,\nonumber
\ee
one gets
\be\label{rhoin}
\phi(r,t)= \int_0^{+\infty}  \rmd\rho\,\mu(\rho)\ w(t, 2\sqrt{r\rho})\,,
\ee
where we defined the measure
\be
\mu(\rho)\equiv \frac{\rme^{-\rho}}{\sqrt{\pi\rho}}\,,\qquad \int_0^{+\infty}\rmd\rho\,\mu(\rho)=1\,.\label{mu}
\ee
Now all the $(t,r)$ dependence is in $w$ and since $\phi$ solves the diffusion equation in these variables, so must $w$ at least upon integration. To prove it, one notices that 
$w$ solves, say, the Laplace equation with respect to the $(t,\s)$ variables but, on the other hand, inside the integral a second derivative with respect to $\s$ can be replaced by a first derivative with respect to $r$. Then, the Laplace equation is equivalent to the heat equation (if $w$ obeys the wave equation, one will get the heat equation with opposite coefficient). In fact,
\ba
\int_0^{+\infty} \rmd\rho\,\mu(\rho)\ \partial_r w(t, 2\sqrt{r\rho})&=& \int_0^{+\infty}  \rmd\rho\,\frac{\rme^{-\rho}}{\sqrt{\pi r}}\ \partial_\s w(t, 2\sqrt{r\rho})\cr
&&\cr
&=& -\left.\frac{\rme^{-\rho}}{\sqrt{\pi r}}\partial_\s w \right|^{+\infty}_0\cr
&&\cr
&&+\int_0^{+\infty}  \rmd\rho\,\frac{\rme^{-\rho}}{\sqrt{\pi r}}\  \partial_\rho \partial_\s w(t, 2\sqrt{r\rho})\cr
&&\cr
&=& \int_0^{+\infty}  \rmd\rho\,\mu(\rho)\ \partial^2_\s w(t, 2\sqrt{r\rho})\cr
&&\cr
&=& -\int_0^{+\infty}  \rmd\rho\,\mu(\rho)\ \partial^2_t w(t, 2\sqrt{r\rho})\,,
\ea
where $\s =2\sqrt{\rho r}$ and in the last equality we used eq.~\Eq{eqb}.

Moreover, the $r$ and $\rho$ dependence of $w$ is symmetric, so that a logarithmic derivative in $r$ is also equal to a logarithmic derivative with respect to $\rho$:
\be
r\p_r w = \rho\,\p_\rho w\,.
\ee
This implies that, when integrated with measure $\mu(\rho)$, $w$ obeys a diffusion equation in $\rho$ and $t$ like eq.~\Eq{difeq2}, which differs from eq.~\Eq{difeq} simply by how the extra coordinate transforms: rescaling in the first case (eq.~\Eq{resc}), translating in the second.

Therefore, we conclude that
\begin{quote}
\emph{$\phi(r,t)$ can be written as an integral with measure $\mu(\rho)\rmd\rho$, eq.~\Eq{mu}, times a function $w(t, 2\sqrt{r\rho})$ which, upon integration, is a solution of the diffusion equation in $\rho$ and $t$.}
\end{quote}
In section \ref{diem} we outlined the diffusion equation method as conceived in previous papers. There, we identified spacetime solutions to the diffusion equation in $r$ frozen at a particular value $r=r_*$ \cite{cuta4} with solutions of the dynamical system. The integral representation \Eq{rhoin} brings a different and, as we shall see, more fertile perspective. The parameter $r$ is a rescaling of the argument of a solution of the heat equation in $\rho$, whose value is determined by the system dynamics.

%%%%%%%%%%%%%%%%%%%%%%%%%%%%%%%%%%%%%%%%%%%%%%%%%%%%%%%%%%%%%%%%%%%%%%%%%%%%%%%%%%%%%%%%%%%%%%%%%%%%%%%%%%%%%%%%%%%%%%%%%%%%%%%%%%%%%%%%%%%%%%%%%%%%%%%%%%%%%%%%%%%%%%%%%%%%%%%%%%%%%%%%%%%%%%%%

\section{Diffusion equation in string theory}\label{sads}

In a series of papers, we developed the diffusion equation method as a tool to solve the spacetime effective equation of motion of the string tachyon and the $p$-adic string. In most cases these solutions were approximate but the level of approximation was good and under control. Some results, such as the brane tension ratio of \cite{cuta5}, were perhaps impressive. Regardless the encouraging positivity of these achievements, the situation is unsatisfactory because we do not have an explanation of the method within OSFT. Without such explanation, the method would be just a fortunate, sometimes miraculous, trick. In fact, it applies to the lowest-truncation-level equation of motion of just one field of the string spectrum, and the obtained solutions are not exact. There is no obvious reason why, in such a crude scenario, one would obtain correctly all the qualitative and most of the quantitative features of tachyon condensation. Moreover, the main drawback of the method is that it provides no existence condition for the solutions of the non-local scalar equation, also because there is no systematic way to choose the initial field configuration. Summarizing, it is desirable to answer the following questions:
\begin{itemize}
\item[\textbf{(Q1)}] Can we justify the status of the diffusion equation method within OSFT?
\item[\textbf{(Q2)}] How to choose the initial conditions of the diffusion equation? Why, and in which sense, is the kink solution more accurate than the one with wild oscillations?
\item[\textbf{(Q3)}] Can the spacetime-based method yield information which presently has not yet been extracted from the full theory? Do we expect the kink solution to have a counterpart in the full theory and, if so, of which form?
\end{itemize}
We shall now partly fill these gaps.

Having revisited some results in the worldsheet and effective spacetime formulations, we are in a position to interpret the diffusion equation in terms of the former. The starting point is to observe, for instance in the Polyakov action \cite{Pol98,Zwi09}, that a conformal transformation $\O^2(z)$ of the worldsheet metric can be regarded also as a conformal rescaling $\O^2(X)$ of the $D$-dimensional target metric $\eta_{\mu\nu}$ (the converse is not true).

This statement filters down to the effective theory as follows. Consider a $D$-dimensional metric $\bar{g}_{\mu\nu}$ and a conformal transformation $g_{\mu\nu}\equiv\O^{-2}\bar{g}_{\mu\nu}$, where $\O=\O(x)$ is a function of the coordinates. We also define the vector $\O_\mu\equiv \p_\mu\ln\O$. Let $\phi$ be a massless scalar field which obeys the free Klein--Gordon equation 
\be\label{mkg}
\bar\Box\phi=0
\ee
in the $\bar{g}$ frame.
% $\bp^\mu\equiv \bar{g}^{\mu\nu}\p_\nu=\O^{-1}\bp^\mu$,,
%Metric and affine connection:
%\ba
%\G^\l_{\mu\nu} &=& \bG^\l_{\mu\nu}-\left[\delta^\l_{(\mu}O_{\nu)}-\frac12 \bg_{\mu\nu}O^\l\right]
%\ea
Then, in the other frame
\be\label{D}
\O^2\bar\Box\phi = \left[\Box+(D-2)\O_\mu\p^\mu\right]\phi=0\,.
\ee
The above equation is nothing but the diffusion equation \Eq{difeq} as soon as one denotes the Lie derivative along the vector $\O^\mu$ as $\p_r=(D-2)\cL_{\vec{\O}}$.

The choice of metric $g_{\mu\nu}$ and initial conditions will determine which class of functions $\O(x)$ realizes the conformal transformation dual to the diffusion process associated with the solution $\phi(r,x)$. For instance, let us take the kink eq.~\Eq{erf} solving the diffusion equation $(\p_x^2-\p_r)\phi=0$. A quick check shows that the resulting conformal factor is
\be
\O(x)=\O_0\exp\left[-\frac{x^2}{4(D-2)r}\right]\,,
\ee
where $\O_0$ is an arbitrary constant.

%Also because we require a solution of the diffusion equation to solve a non-local Klein--Gordon equation with non-linear potential, 
The conformal transformation can be seen as one going from zero\footnote{One can always redefine the scalar field so that a mass term be reabsorbed. Given a constant $\b$, the field $\bar\phi=\rme^{-\b r}\phi$ obeys the massive diffusion equation $(\B+\b+\p_r)\bar\phi=0$.} to finite momentum. As we have recalled, in the CFT state space this can be understood as a \emph{reparametrization of the vacuum. But the diffusion equation \Eq{difeq} (whose CFT counterpart are eqs.~\Eq{bide} and \Eq{difeqgh}) is a background-dependent way to implement it}. Therefore, in the effective theory one has to find the equivalent of the operator $U$ of section \ref{gau} realizing a large/singular gauge transformation.

Recently, the vacuum solution \Eq{vacs} was written in a very appealing form, that is, as an integral over wedge states \cite{ES09}:
\be\label{vsoin}
\Psi=\int_0^{+\infty} \rmd r \rme^{-r} P W_{r}\,,
\ee
where $P=c+cL^+B^+c/4$. For non-vacuum solutions, the form of the operator $P$ will be different. We are now able to collect evidence that a large class of exact OSFT solutions admit an integral representation where they are expanded on a basis of diffusing states (surface states). 

This `large class' contains at least the exact analogues of the known spacetime diffusing solutions (bosonic marginal deformations, supersymmetric marginal deformations, supersymmetric kink configurations), and it is likely to be much wider. The point is that eq.~\Eq{rhoin} is the spacetime counterpart of eq.~\Eq{vsoin}. Since (a) diffusing solutions of the target effective theory, bosonic and supersymmetric, have all the same structure as the known full solutions, (b) they all admit the integral representation \Eq{rhoin}, and (c) they include marginal deformations and non-trivial configurations such as kinks (full brane decay), then also the CFT counterparts of the wild oscillations (so far known only as a perturbative series in the bosonic theory) and the kink (so far unknown) should be of the form \Eq{vsoin}. This conclusion is supported also by the discussion in section \ref{gau} on the bosonic string.

In fact, \emph{the diffusion equation can be regarded as a change of gauge which simplifies the problem} and recasts the effective dynamics of OSFT in a gauge convenient form. In particular, in \cite{cuta5} we started from a solution around the tachyon vacuum, i.e., a distributional constant solution $\phi={\rm sgn}$, representing the minima of the effective double-well potential. This is tantamount to asking that in the gauge frame $\bar g$ the solution be constant (eq.~\Eq{D}) and extremizes the effective potential, so that the equation of motion \Eq{eom} is the free equation. In this frame the solution plays the same role of a projector, as it is idempotent ($\Psi*\Psi=\Psi$) and closed ($Q\Psi=0$) in the sense of eq.~\Eq{mkg}. Equations \Eq{wed} and \Eq{surfsd} are nothing but the universal version of the statement that solutions of the diffusion equation are of the form eq.~\Eq{solde} and one can identify surface states (eventually with insertions) as such solutions, and the initial condition in eq.~\Eq{pip2} which evolves via eq.~\Eq{rhoin} as a projector which `evolves' via eq.~\Eq{vsoin}. The conformal mappings describing the sliver state are singular but the state itself is well-defined \cite{RSZ3}. In the background-dependent framework of the diffusing system, the initial condition in diffusion time may be a discontinuous distribution, but the diffusion flow smears it to a smooth spacetime function, eq.~\Eq{rhoin}.

This should answer question {\bf (Q1)}. The diffusion equation method works because it is recognized to be an implementation at spacetime level of the gauge (reparametrization) freedom of the full theory. Therefore, spacetime solutions inherit the diffusing property of the exact ones, as long as the initial conditions are chosen correctly. This is unexpected. One of the most beautiful achievements in string theory is that non-trivial solutions to the equation of motion are built starting from a non-normalizable state living in an `enlarged' Hilbert space and performing a large or singular gauge transformation on this state, which `drags' it into the physical Hilbert space. The non-normalizable state is a trivial formal solution, the ending point is a non-trivial physical solutions. Here we find that the same construction applies to \emph{effective spacetime solutions}. In general relativity or quantum field theory we do not enjoy of such inheritance property: to find a solution entails a gauge choice (of the metric, of the frame, etc.) which spoils the symmetries of the theory (diffeomorphism and Lorentz invariance). Here, on the other hand, we start with a theory endowed with a certain symmetry group. We make some approximations, choose frames and CFT's (backgrounds), employ a mysterious diffusion equation, and so on. We end up with solutions which common sense would label as qualitative at best. They are not, because the `diffusing' character of the full theory propagates down to them. This has very concrete consequences, partly explored in previous papers and partly below. In this sense, the most logical and appealing way to present the method would have been first to understand its interpretation, and then support it via the results of \cite{roll,cuta3,cuta4,cuta5}.

An answer to question {\bf (Q3)} is also within reach. We already have two concrete examples of solutions not yet constructed in the full theory: a supersymmetric OSFT profile with wild oscillations and a kink solution obeying Sen's descent relation. The last is important also for another reason. The conjecture that general exact solutions are superpositions of surface states is not new, but here we formulate it upon the integral form eq.~\Eq{vsoin} and include  non-marginal deformations explicitly, while noticing the existence of examples (wild oscillations, kink) supporting the claim.

%%%%%%%%%%%%%%%%%%%%%%%%%%%%%%%%%%%%%%%%%%%%%%%%%%%%%%%%%%%%%%%%%%%%%%%%%%%%%%%%%%%%%%%%%%%%%%%%

\subsection{Why spacetime diffusing solutions are approximate (but good)}\label{good}

Diffusing solutions of the effective lowest-order theory do not encode all the information of a full solution but they capture their main behaviour according to their accuracy. This was verified both for the rolling tachyon with marginal deformations in bosonic and susy OSFT \cite{roll} and the kink solution of susy OSFT \cite{cuta4,cuta5}. In the first case, the solution reproduced all the qualitative features of the wild oscillations, the coefficients of the series representation being all very close to those obtained with perturbative techniques. In the second case, the solution was global and very accurate, and it was shown to realize a brane decay, the brane tension ratio being close to the expected value at $1\%$ level. These solutions are lowest-level in the truncation scheme and approximate, yet they describe the tachyon dynamics quantitatively in agreement with independent results. 

From the usual formulation of the diffusion method of section \ref{diem}, it would be natural to explain why spacetime solutions are approximate by interpreting them as `semiclassical' solutions peaked at one surface state in the continuum basis. Then, one could try to improve a solution by taking finite or infinite superpositions of copies of the solution with different $r$'s. However, the reinterpretation of section \ref{integr} via eq.~\Eq{rhoin} demonstrates that all diffusing spacetime solutions can already be written as integrals of formal solutions of the diffusion equation times an appropriate weight function. Thus, any attempt to reduce the global error by considering finite or infinite superpositions of solutions is bound to fail. We checked it explicitly. Even if these `integrated solutions' are not of much use in string theory, below we report these results in detail because they will incidentally indicate a route towards solutions of other non-local models such as the $p$-adic string.

Having excluded the `semiclassical' interpretation of spacetime approximate solutions, the most obvious source of inaccuracy is level truncation. This claim is not trivial. On one hand, the level-0 approximation is good, at least for marginal deformations and brane decay configurations, only because non-locality (off-shell potential) is taken into account. On the other hand, the diffusion equation method yields approximate solutions of the truncated effective theory but its roots go beyond level truncation, deep into the structure of the full theory (eqs.~\Eq{bide} and \Eq{vsoin}). This is why all qualitative and most of the quantitative features of the full theory of tachyon condensation are captured by diffusing spacetime solutions.

Question {\bf (Q2)} has also been answered in part. The diffusion equation is nothing but a reparametrization of a solution which is trivial in the distributional sense (the initial condition), so non-trivial initial profiles should be illegal. If the initial conditions are close enough to the analogue of a `trivial state', then they should still be able to catch some qualitative features. A comparative example is the kink solution versus wild oscillations (below analyzed in detail). Also, solutions of the diffusion equation are already superpositions of diffusing states, so it is not useful to consider such superpositions. The next section provides explicit examples.

Interestingly, there may be also another explanation for the approximate nature of spacetime diffusing solutions. Before the discovery of Schnabl's gauge, a method for finding normalizable solutions was proposed \cite{GRSZ1}. There, VSFT is regularized, so that the regularized sliver state is an approximate solution of the theory when the regulator is kept finite, and it reduces to the sliver projector when the latter is removed. In particular, one defines a deformed Siegel gauge (or deformed CFT) with kinetic operator ${\cal Q}=c_0(1 + a^{-1}L_0)$. When $a\to 0$, ${\cal Q}$ tends to the BRST operator, while when $a\to\infty$ it is the pure ghost operator of VSFT. In \cite{Sc02b,Ok03b} it was shown that there exists a unique (regularized) gauge invariant surface state in (regularized) bosonic VSFT, and that upon removing the regulator this state is the butterfly and solves VSFT exactly. If the regulator $a$ is kept finite, then the `deformed butterfly' is an approximate solution of the equation of motion (it is a projector only at leading order in a regulator expansion). A generalization to other projectors was studied in \cite{Yan04}.

The spacetime diffusing solutions could be regarded also as configurations with a finite regulator. Then, the coefficient $a^{-1}$ in front of the kinetic operator can be reabsorbed in a coordinate redefinition, which in turn can be seen as a rescaling of the perturbative tachyon mass. Extending the same philosophy to the supersymmetric theory, this would explain why the most accurate kink solution of the OSFT lowest-level action with non-local potential (eq.~\Eq{susa}) featured a value $r_*$ different from that of OSFT (this can be readjusted by rescaling the coordinate and then the mass) \cite{cuta5}. On the other hand, the solution of the simplified system with approximate non-locality (local potential, eq.~\Eq{a1}) has the usual value of the mass \cite{cuta5}. In either case, the expected brane tension ratio is reproduced, so at this stage it is not clear whether the finite regulator picture is useful or not.

\section{Applications to OSFT and non-local theories}\label{appl}

We would like to see whether and how the results of the previous section affect the construction of spacetime solutions. We have already argued that the diffusion equation method is a spacetime implementation of the gauge freedom of the full theory, thus explaining its physical origin. This also enables us to better select the initial conditions or combinations of diffusing states which will produce a solution of the equation of motion; sections \ref{wild}, \ref{fsup} and part of \ref{isup} are devoted to this task. The rest of \ref{isup} and section \ref{pol} discuss the general structure and some examples of approximate solutions of string-inspired non-local models.

%%%%%%%%%%%%%%%%%%%%%%%%%%%%%%%%%%%%%%%%%%%%%%%%%%%%%%%%%%%%%%%%%%%%%%%%%%%%%%%%%%%%%%%%%%%%%%%%

\subsection{Wild oscillations}\label{wild}

As an application of formula \Eq{rhoin}, one can easily recover eq.~\Eq{nloc3x} starting from the initial condition
\be\label{badic}
\phi(0,t)=\tfrac32\,{\rm sech}(t)\,.
\ee
For brevity, we have rescaled time by a factor $\sqrt{2}$ which can be restored at the end of the calculation. An analytic function  $f(t+\rmi\s)$ whose real part coincides with $\psi(0,t)$ at $\s=0$ is, by construction,
\ba
f(t+\rmi\s)&=& \frac32\frac{1}{\cosh(t+\rmi\s)}\cr
&&\cr
&=&\frac{3\cos(\s)\cosh(t)}{\cos(2\s)+\cosh(2t)}-\rmi\frac{3\sin(\s)\sinh(t)}{\cos(2\s)+\cosh(2t)}\,.
\ea
When $t<0$ we get
\ba
\phi(r,t)&=&\frac{3}{2}\int_0^{+\infty}\rmd\rho\,\frac{\rme^{-\rho}}{\sqrt{\pi\rho}}\ {\rm Re}\left[\frac{1}{\cosh(t+2\rmi\sqrt{r\rho})}\right]\cr
&&\cr
&=& 3\int_0^{+\infty}\rmd\rho\,\frac{\rme^{-\rho}}{\sqrt{\pi\rho}}\ {\rm Re}\left[\sum_{n=0}^\infty (-1)^n\rme^{(2n+1)(t+\rmi 2\sqrt{r\rho})}\right]\cr
&&\cr
&=& 3\sum_{n=0}^{+\infty}(-1)^n \rme^{-r(2n+1)^2}\rme^{(2n+1)t},
\ea
which indeed coincides with eq.~\Eq{nloc3x} under rescaling $t\to t/\sqrt{2}$. In the above equation, the domain $t<0$ was chosen to justify the expansion of the hyperbolic secant and the subsequent term-by-term integration. Once integrated, the convergence domain of the series extends to the whole $t$ axis.

The bosonic and susy solutions with wild oscillations of \cite{roll} were not global solutions, although they did capture the qualitative behaviour of the well-known solutions in series representation \cite{MZ,KORZ,CST,FGN2}. The reason is now clear: even if the initial condition \Eq{badic} is a solution of the susy equation of motion with $r=0$, this corresponds to case (B2) of section \ref{diem}, which does not correspond to a free field solution (i.e., one extremizing the tachyon potential). This might suggest that case (B2) never leads to exact solutions, as one could have already noticed from the discussion of section \ref{sads}. A way to find global solutions with continuous initial conditions is to generalize the free equation \Eq{mkg} in the $\bar g$ frame with an interaction or a source term. Only restrictions on the source term will allow one to solve the dynamical problem concretely. Let the inhomogeneous diffusion equation be $(\B+\p_r)\phi=\B f$, where $f(x)$ is some function. The non-local exponential operator acts on $\phi$ as a translation along the $r$ direction, plus an extra contribution:
\be
\rme^{s\B}\phi(r,x)=\phi(r-s,x)+\left(\rme^{s\B}-1\right)\!f\,.
\ee
If $f$ is a polynomial or an eigenstate of $\B$, the non-local term in the right-hand side can be computed explicitly.

Here we will not consider the interesting consequences of this modification of the diffusion equation recipe. For the time being, we stress that case (B2) is still allowable from a phenomenological point of view, at least for the bosonic rolling tachyon with marginal deformations. Since this is the only non-trivial solution we know in \emph{both} the full bosonic theory and the effective spacetime picture, it is worth collecting sparse results in the literature on the coefficients $c_n$ of the bosonic series eq.~\Eq{solmd} and draw an explicit comparison of the numerical values obtained with perturbative techniques \cite{CST,FGN2}, the diffusion equation method \cite{roll,FGN} and the exact result in the full theory \cite{KORZ}. These are summarized in table \ref{tab1}. The coefficient $c_2=2^6/3^{11/2}\approx 0.152$ is the same in all cases. Errors of order $100\%$ are still acceptable as the coefficients have the same order of magnitude and are exponentially suppressed (in fact, a better estimate of the error is on the logarithmic coefficients, $\tilde\Delta_n$ in figure \ref{fig1}). With the only exception of level-2 $c_5$, all the other coefficients of the diffusing non-perturbative series are much closer to the exact values than those computed with other methods. This is because the level-0 calculation takes non-local effects into account, while higher-level results are obtained on-shell (local models).
\TABLE{%\begin{table*} %[ht]
%\begin{ruledtabular}
\begin{tabular}{|c||l|l|l||l|}\hline\hline
      & {\footnotesize Pert.~$L2$ \cite{CST}} & {\footnotesize Pert.~$L16$ \cite{FGN2}} & {\footnotesize \textbf{Non-pert.~$L0$ \cite{roll,FGN}}} & {\footnotesize Exact \cite{KORZ}} \\ \hline\hline
{\footnotesize $c_3$} & {\footnotesize $2.187 \cdot 10^{-3}$ (1.8\%)} & {\footnotesize $2.342 \cdot 10^{-3}$ (9\%)} & {\footnotesize $\bm{2.139 \cdot 10^{-3}}$ \textbf{(0.4\%)}} & {\footnotesize $2.148 \cdot 10^{-3}$} \\
{\footnotesize $c_4$} & {\footnotesize $3.926 \cdot 10^{-6}$ (50\%)} & {\footnotesize $4.844 \cdot 10^{-6}$ (85\%)} & {\footnotesize $\bm{3.297 \cdot 10^{-6}}$ \textbf{(26\%)}} & {\footnotesize $2.619 \cdot 10^{-6}$} \\
{\footnotesize $c_5$} & {\footnotesize $4.941 \cdot 10^{-10}$ (77\%)} & {\footnotesize $5.134 \cdot 10^{-9}$ ($\gg$)} & {\footnotesize $\bm{5.876 \cdot 10^{-10}}$ \textbf{(111\%)}} & {\footnotesize $2.791 \cdot 10^{-10}$} \\
{\footnotesize $c_6$} & {\footnotesize $-6.323\cdot 10^{-12}$ ($\gg$)} & {\footnotesize $7.357\cdot 10^{-13}$ ($\gg$)} & {\footnotesize $\bm{1.240 \cdot 10^{-14}}$ \textbf{(343\%)}} & {\footnotesize $2.801 \cdot 10^{-15}$} \\
{\footnotesize $c_7$} & {\footnotesize --}      & {\footnotesize --}   & {\footnotesize $\bm{3.136 \cdot 10^{-20}}$ \textbf{($\bm{\gg}$)}} & {\footnotesize $2.729 \cdot 10^{-21}$}\\
%\begin{tabular}{|l||c|c|c||c|}
 % & Pert. level-2 \cite{CST} & Pert. level-16 \cite{FGN2} & Pert. level-2 (exact 4-tachyon amp.) \cite{FGN2} & \textbf{Nonpert. \cite{roll,FGN}} & Exact \cite{KORZ} \\ \hline\hline
%$c_3$ & $2.187 \cdot 10^{-3}$ (1.8\%) & $2.342 \cdot 10^{-3}$ (9\%)) & $2.415 \cdot 10^{-3}$ (12\%) & $\bm{2.139 \cdot 10^{-3}}$ \textbf{(0.4\%)} & $2.148 \cdot 10^{-3}$ \\
%$c_4$ & $3.926 \cdot 10^{-6}$ (50\%) & $4.844 \cdot 10^{-6}$ (85\%) & $5.349 \cdot 10^{-6}$ (104\%) & $\bm{3.297 \cdot 10^{-6}}$ \textbf{(26\%)} & $2.619 \cdot 10^{-6}$ \\
%$c_5$ & $4.941 \cdot 10^{-10}$ (77\%) & $5.134 \cdot 10^{-9}$ & $2.065 \cdot 10^{-9}$ & $\bm{5.876 \cdot 10^{-10}}$ \textbf{(111\%)} & $2.791 \cdot 10^{-10}$\\
%$c_6$ & $-6.323\cdot 10^{-12}$ & $7.357\cdot 10^{-13}$ & --                    & $\bm{1.240 \cdot 10^{-14}}$ \textbf{(343\%)} & $2.801 \cdot 10^{-15}$\\
%$c_7$ & --                     & --   & --             & $\bm{3.136 \cdot 10^{-20}}$ & $2.729 \cdot 10^{-21}$\\
\hline\hline\end{tabular}%\end{ruledtabular}
\caption{\label{tab1} Comparison of the coefficients $c_n$ of the exact rolling tachyon solution of \cite{KORZ} with the perturbative calculations at level $L$ \cite{CST,FGN2} and the non-perturbative solution found with the diffusion equation method \cite{roll,FGN}. The percentage in brackets denotes the deviation $\Delta_n\equiv|c_n-c_n^{\rm exact}|/|c_n^{\rm exact}|$ from the exact value. When indicated as $\gg$, this error is about or larger than one order of magnitude.}
}%\end{table*}

The coefficients of the exact solution are given in integral form, and numerically up to $n=7$, in \cite{KORZ}. We checked that they can be well described by the non-linear fit
\be\label{cne}
c_n^{\rm exact} \approx 1.069n\,\rme^{1.032n-1.154n^2}\,,
\ee
%\be
%c_n^{\rm exact} \sim 2.7 \cdot 10^{-n(n-1)/2}\,.
%\ee
while the coefficients \Eq{psix} of the level-0 diffusing solution are
\be\label{psix2}
c_n\approx 1.622n\,\rme^{0.563 n-1.046 n^2}\,.
\ee
The coefficients ratio is, respectively,
\be\label{coex}
\frac{c_{n+1}^{\rm exact}}{c_n^{\rm exact}}\approx 0.885\rme^{-2.309 n}\left(1+\frac1n\right)\,,
\ee
%\be\label{coex}
%\frac{c_{n+1}^{\rm exact}}{c_n^{\rm exact}}\sim \rme^{-2.303 n}\,.
%\ee
and
\be\label{rat}
\frac{c_{n+1}}{c_n}\approx 0.617 \rme^{-2.093 n}\left(1+\frac1n\right)\,.
\ee
Both numbers in eq.~\Eq{rat} are slightly smaller than those in eq.~\Eq{coex}, which is why the coefficients $c_n$ of the diffusing solution are greater than the exact ones (see table \ref{tab1} and figure \ref{fig1}). At large $t$, the time exponentials start to dominate and amplify the error on the coefficients. In this sense, the diffusing solution of \cite{roll,FGN} was claimed to be a good solution for not too large $t$ \cite{roll}. However, eqs.~\Eq{coex} and \Eq{rat} show that the diffusing solution well captures the behaviour of the exact solution at all times, not only asymptotically; see figure \ref{fig1}.
\FIGURE{
\includegraphics[width=9.6cm]{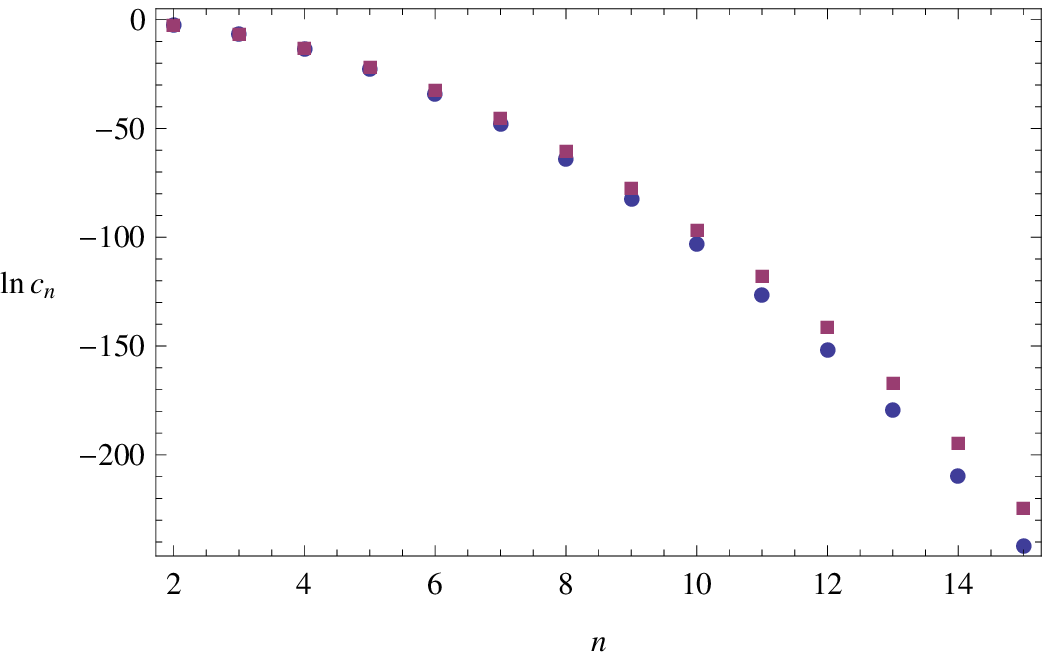}
\includegraphics[width=9.6cm]{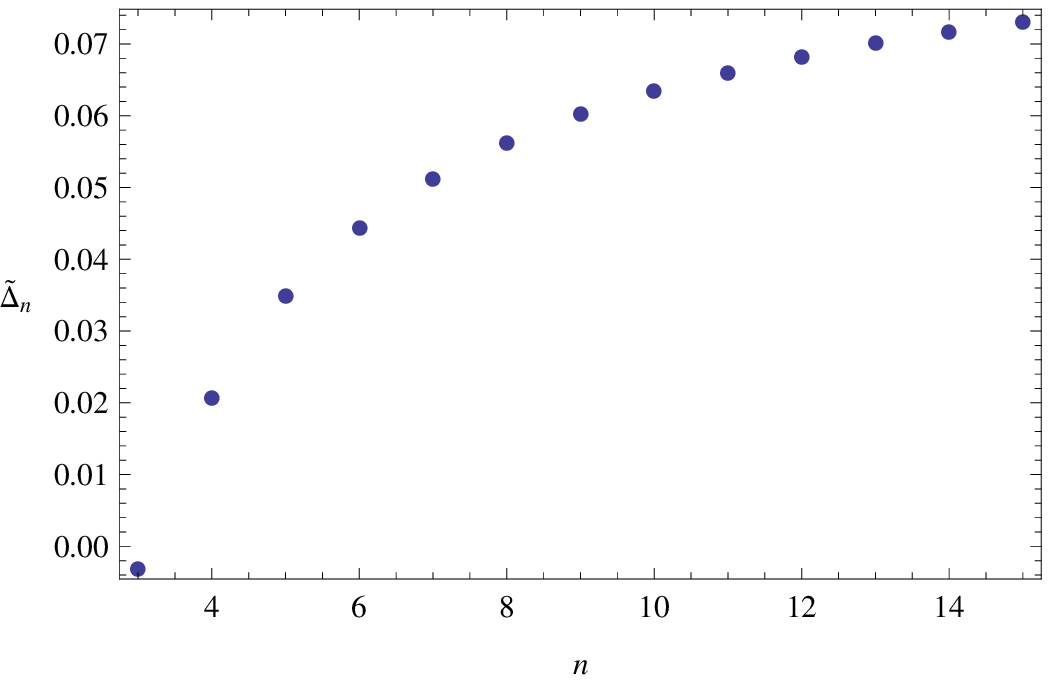}
\caption{\label{fig1}
Top panel: the coefficients $\ln c_n^{\rm exact}$ (dots) and $\ln c_n$ (squares) of the exact and diffusing series, respectively. Bottom panel: $\tilde\Delta_n\equiv |\ln c_n-\ln c_n^{\rm exact}|/|\ln c_n^{\rm exact}|$.}}

Since the $n$-th element of the level-0 non-perturbative series obeys the diffusion equation individually,
the spacetime effective solution \Eq{solmd} can be already considered a `superposition of wedge states,' as is clear from the structure of the exact coefficients (see \cite{KORZ}, eq.~(3.3)). The differences, small as they are, between eqs.~\Eq{coex} and \Eq{rat} can be ascribed mainly to the truncation expansion, absent in the first case and lowest-order in the second.

To improve the solution \Eq{solmd} one could leave $r_*$ arbitrary and see what is the best fit for the exact coefficients, or what is $r$ such that one obtains one of the coefficients in eq.~\Eq{cne}. In either case, the resulting $r$ would not be the same as eq.~\Eq{rst}, in accordance with the finite-regulator interpretation. This situation is remindful of the kink solutions of \cite{cuta5}. There, and for the potential in eq.~\Eq{susa}, one could either obtain an approximate solution for $r=r_*$ or another, more accurate, with different $r$. The latter is a solution of an equation of motion formally identical with the OSFT effective equation, but with other values of the constants. However, the same type of solution (with $r\neq r_*$) might actually arise from a non-local diffusion equation of the form $\B\phi(r',x)+\p_r\phi(r,x)=0$, where $r\neq r'$. Other possibilities, such as an inhomogeneous diffusion equation, were discussed in section \ref{sads}.

Another option is to consider the fact that the $r,n$ dependence of the diffusing series coefficients is simple for all $n$. Each coefficient $c_n$ in the exact series is given by a multiple integration of operators along the length of the strip corresponding to the $n$-th wedge state. The $n$-dependence from the length of the wedge state is therefore difficult to assess, and in general some approximation should be employed to obtain a simple expression like eq.~\Eq{psix}.\footnote{For instance, the mean value theorem for integration applied to eq.~(3.34) of \cite{KORZ} does work for $n\leq 5$ ($n\leq4$ in the notation of \cite{KORZ}). Surprisingly, the point in the interval $[0,1]$ which realizes this approximation is about the same for all $n\leq 5$ (and equal to $t_n=0.375$). This way the first coefficients acquire a simpler $n$-dependence. As $n$ increases, however, the error propagates and the approximation breaks down.} Therefore it would be natural to replace $r_*\to r_n$ and make a fit which solves the OSFT effective equation of motion. However, the discussion in section \ref{good} suggests that naive superpositions of a spacetime solution with different $r$'s would not yield improved solutions. It may be instructive to show this in a particular instance.

%%%%%%%%%%%%%%%%%%%%%%%%%%%%%%%%%%%%%%%%%%%%%%%%%%%%%%%%%%%%%%%%%%%%%%%%%%%%%%%%%%%%%%%%%%%%%%%%

\subsection{Finite superposition of solutions}\label{fsup}

A new kink-type candidate solution could be made of the sum of a number of copies of the kink eq.~\Eq{erf}, each with coefficient $r_n\neq r_l$. This might increase the accuracy of the result with respect to the simplest case $r_n=r_l$, $\forall\, l,n$. Let us take the superposition of just two one-dimensional kinks with same asymptotics as eq.~\Eq{erf},
\be\label{2erf}
\phi=C\,\erf\left(\frac{x}{\sqrt{4r_1}}\right)+(1-C)\,\erf\left(\frac{x}{\sqrt{4r_2}}\right)\,,
\ee
where $C$ is a constant. For illustrative purposes and without loss of generality, we are interested in a simplified version of eq.~\Eq{susa} \cite{AJK} where the scalar potential is a pure power without non-local insertions:
\be\label{a1}
\left(\p_x^2+\frac12\right)\rme^{-2r_*\p_x^2}\phi=\s\phi^3\,.
\ee
For the kink solution, this approximation was shown to lead to the same dynamics of eq.~\Eq{susa} \cite{cuta5}.

The asymptotics at $x=\infty$ fixes the coupling $\s=1/2$ (or, alternatively, the normalization of the solution for a given $\s$). Moreover, since the leading term in a $x\sim 0$ expansion of the right-hand side of eq.~\Eq{a1} is cubic, the $O(x)$ term in the left-hand side must vanish. This fixes the coefficient $C=C(r_1,r_2,r_*)$.
%\be
%C=\frac{1-r_2+2r_*}{1-r_2+2r_*-(1-r_1+2r_*)\left(\frac{r_2-2r_*}{r_1-2r_*}\right)^{3/2}}\,.
%\ee
Setting, e.g., $r_1=1.5$, there remains only one free parameter to be tuned in order for the error on the equation of motion to be minimized. The latter is defined as 
\be\label{DX}
\Delta_{\rm max}\equiv \mathop{\rm sup}_{x}\Delta(x)\equiv \mathop{\rm sup}_{x}\left|\frac{{\rm l.h.s.}-{\rm r.h.s.}}{{\rm scale}}\right|\,,
\ee
where l.h.s. and r.h.s. are, respectively, the left- and right-hand side of eq.~\Eq{a1}, and the denominator is some characteristic scale of the solution. A typical choice is ${\rm l.h.s.}+{\rm r.h.s.}$, but others are possible and do not change much the error estimate (see \cite{roll,cuta2,cuta5} for details). One can show that values around $r_2= 1.3$ minimize the error to $\Delta_{\rm max}\approx 1.4\%$. This is only slightly smaller than the error for the single-kink solution of the equation of motion with the same values of the parameters $r_*$, $m$ and $\s$, which is $\Delta_{\rm max}\approx 1.5\%$ (it can be calculated directly on the second duplication formula of \cite{cuta5}).

Repeating the same procedure for a three-kink solution, we checked that it is possible to fix the parameters $r_{1,2,3}$ and the coefficients of the linear combination so that the error is about $\Delta_{\rm max}\approx 1.4\%$, but not lower. Therefore, the tachyonic kink solution is not improved appreciably by a linear superposition of kinks.

Rather than fixing $r_1$ beforehand, one can set all the free parameters by imposing the coefficients of the coordinate expansion near the origin to vanish, but to no avail. This typically happens in cases where one is not using a complete functional basis to express a solution to the equations of motion, but we have shown that this is not the case. A possibility, which we shall not pursue here, is to consider a finite linear combination of diffusing solutions with different boundary conditions. Another is to take an infinite superposition of kinks.

%%%%%%%%%%%%%%%%%%%%%%%%%%%%%%%%%%%%%%%%%%%%%%%%%%%%%%%%%%%%%%%%%%%%%%%%%%%%%%%%%%%%%%%%%%%%%%%%

\subsection{Integrated solutions of non-local models}\label{isup}

Let $\phi(r,x)$ be an approximate or exact solution of both the diffusion equation and the non-local equation of motion of a given model. We define as the `integrated solution' the function
\be\label{ints}
\psi=\int_I\rmd r\, \mu(r)\phi(r,x)\,,
\ee
where $I=[a,b]$, $a$ and $b$ are non-negative and $\mu$ is a one-dimensional measure weight such that 
\be
\int_I\rmd r\, \mu(r)=1\,,
\ee
in order for $\psi$ to have the same normalization as $\phi$.

There is a caveat regarding the integration interval $[a,b]$. Because the heat equation is a conformally transformed free Klein--Gordon equation, all its (one-dimensional) solutions depend on the argument $y\equiv x^2/(4r)$. Every value of $y$ can be achieved by any other value of the space(time) coordinate under a suitable rescaling in $r$. In particular, the points $y=0$ and $y=\infty$ are degenerate as they correspond to two different asymptotics: one in 
space(time) ($x=0$ and $x=\infty$, respectively) and one in the extra direction ($r=\infty$ and $r=0$). Therefore, if one integrates the solution on the whole positive real axis one may encounter unwanted singularities in the solution or its derivatives.

The following proposition (valid in one dimension but easily extendable to the general case) holds which prevents this to happen. All $C^\infty$ integrated solutions \Eq{ints} which can be analytically continued on the whole real axis are of the form
\be\label{intsol}
\psi=\int_0^{+\infty}\rmd r\,\mu(r)\phi(r,y)\,,
\ee
where
\be\label{mur}
\mu^{(n)}(0)=0\,,\qquad n\in \mathbb{N}\,,
\ee
and the superscript $(n)$ denotes the $n$-th derivative. For instance, weights of the form $\mu\sim r^n\rme^{-r}$ produce a discontinuity in $\psi^{(n+2)}$, since $\mu^{(n)}(0)\neq 0$. On the other hand, there exist weights which respect the singularity-free and normalization requirements. Examples are (all $l,n$ positive integers) 
\ba
\mu(r)&\propto&\frac{\rme^{-1/r^n}}{r^2}\,,\qquad
\mu(r)\propto (r^n+r^{-n})^{-l}\,,\\
\mu(r)&=& \frac{4n}{\pi r_0}\left(\frac{r}{r_0}\right)^{n-1}\frac{1}{[(r/r_0)^n+(r_0/r)^n]^2}\,,\label{r0}
\ea
where $I=\mathbb{R}^+$ and $r_0$ is a scale. The last measure has two interesting properties: for $n=1$ it is invariant under the inversion $r\to 1/r$, and in the large $n$ limit it tends to $\mu\sim \delta(r-r_0)$.

As anticipated, the integrated kink (which is still a kink) is no longer a solution of the effective equation of motion with potential $\psi^3$, for any of the above measures. %(the minimum error we were able to obtain was about $8\%$). 
This is true also for eq.~\Eq{susa} (non-local potential).\footnote{To show it, one has to employ the formula
\be
\erf\left(\frac{x}{\sqrt{4r_1}}\right)\erf\left(\frac{x}{\sqrt{4r_2}}\right)\approx 1-\rme^{-\frac{x^2}{\pi\sqrt{r_1r_2}}}\,,
\ee
which can be obtained in a way similar to the duplication formul\ae\ of \cite{cuta5} and is valid also upon integration in $r_1$ and $r_2$. The integration intervals $I_{1,2}=[a,+\infty)$ are chosen so that $a>2r_*$ and both sides of eq.~\Eq{susa} (which have different signs in the non-local exponents, and thus require a limitation of the domains) are well-defined.} 

It is instructive to draw a comparison between non-integrated and integrated solutions for another model, a modified $p$-adic equation of motion:
\be\label{pad}
(\rme^{-s\B}-m^2)\psi =\s\psi^n.
\ee
This toy model was often considered in the literature \cite{Jou07,MuN,NuM,AJ,AJV,AJK2,Ver09} as a useful hybrid between the string field tachyon and the $p$-adic string. On one hand, the mass term is strictly non-zero as for the OSFT tachyon, our reason being technical (see below). On the other hand, the equation of motion becomes purely algebraic in the local limit, as for the $p$-adic string.

The parameter $s$ is a constant we will assume to be either $s=+1$ or $s=-1$. Consider the case $n=3$, $s=-1$ and one-dimensional spatial configurations, $\rme^{-s\B}=\rme^{\p_x^2}$. As a characteristic scale of the problem, we take the denominator of eq.~\Eq{DX} to be 1 (the asymptotics of the solution). Then, one can show that eq.~\Eq{erf} is an approximate solution and the error \Eq{DX} is minimized at $\Delta_{\rm max}\lesssim 0.1\%$ for $r\approx 1.78$. %(Fig.~\ref{fig}).
The mass is fixed by the vanishing of the $O(x)$ term in eq.~\Eq{pad}, $m^2=\sqrt{r/(r+s)}\approx 1.511$ (this is the reason why $m\neq 0$), while the normalization of the potential is $\s=1-m^2<0$.

The integrated solution
\be
\psi(x)=\beta\int_0^{+\infty}\rmd r \frac{\rme^{-\beta/r}}{r^2}\, \erf\left(\frac{x}{\sqrt{4r}}\right)=\frac{x}{\sqrt{x^2+4\beta}}\nonumber
\ee
is a kink with same asymptotics as the error function but different slope. This is a candidate solution for the $n=3$, $s=+1$ system. We checked that the error is minimized for $\beta\approx 0.7$ and $m^2\approx 0.62$ ($\s>0$), but the global error is rather high, $\Delta_{\rm max}\lesssim 6\%$. The same result holds for the measure $\mu=\rme^{-\beta/r^2}/r^2$, but not for eq.~\Eq{r0}: an $r_0=0.33$ yields an error of about $2\%$. All these features are valid also when taking a different integration interval $I$ (e.g., $I=[a,+\infty)$ with $a\neq 0$).

The operation of integrating a solution for a given non-local system with chosen $s$ will yield a candidate solution for the same system but with $s\to -s$. In the $p$-adic case this change of sign can be regarded as a switch from the Minkowski to the Euclidean problem (or vice versa). %The same, with little extra work, can be shown to hold for OSFT solutions.%\footnote{This recalls the dichotomy already noticed in the Minkowski solutions of \cite{roll}. There, the approximate solution was in very good agreement with the tachyon equation of motion ($s>0$) while the Euclidean candidate (a smooth lump, $s<0)$ was not a solution.}

%%%%%%%%%%%%%%%%%%%%%%%%%%%%%%%%%%%%%%%%%%%%%%%%%%%%%%%%%%%%%%%%%%%%%%%%%%%%%%%%%%%%%%%%%%%%%%%%

\subsection{Toy models with polynomial potentials}\label{pol}

We conclude with the construction of kink solutions for a family of toy models of the form
\be\label{fv}
\phi(\ga x)=V[\phi(x)]\,,
\ee
where $\ga$ is a constant. The $p$-adic string equation is of this type. On the left-hand side, the rescaling of the coordinate can be achieved by a generic pseudo-differential operator. %Around the origin,
%\be
%f(\ga x)= \sum_n \frac{(\ga x)^n}{n!}
%\ee
Here we limit ourselves to the exponential operator and $\phi=\erf$. Then, around the origin
\be\nonumber
\sqrt{\pi}\phi(\ga x)= 2 \ga x-\frac{2\ga^3 x^3}{3}+\frac{\ga^5 x^5}{5}+O(x^7)\,.
\ee
To match all powers up to $O(x^{l+1})$, the right-hand side of eq.~\Eq{fv} should be of the form
\be
V[\phi]=\sum_{n=0}^l a_n(\ga) \phi^n\,,
\ee
for some set of coefficients $a_n(\ga)$ which can be obtained recursively order by order. In this case, $a_{2n}=0$, $a_1=\ga$, $a_3=-\pi\ga(\ga^2-1)/12$, and so on. The solution at large $x$ is fixed by the polynomial equation
\be
\sum_{n=0}^l a_n(\ga)=1\,,\label{an}
\ee
Up to $O(x^{12})$, one gets an $l=11$ potential and a solution with global error $\Delta_{\rm max}\approx 10^{-5}$ for any of the four allowed values of $\ga\neq 1$ according to eq.~\Eq{an}.

The massive $p$-adic model \Eq{pad} corresponds to the case $l=3$. Equation \Eq{an} gives $\ga=(\sqrt{1+48/\pi}-1)/2\approx 1.517$ (then, $a_3=1-\ga\approx -0.517$) and minimizes the error to $\Delta_{\rm max}\approx 0.1\%$, all numbers in agreement with section \ref{isup}. The error can be lowered to $0.04\%$ by a slightly different value, $\ga=1.511$.

%because only the first two non-vanishing coefficients of the Taylor expansion of the rescaled solution are constrained. At intermediate $x$, in fact, the solution breaks down. An alternative path, valid at any order, is to look for $\ga$ fixed points, i.e., values of $\ga$ which set all powers in the right-hand side to zero for $n\geq \bar n$. For instance, $\ga=0$ and $\ga=1$ are trivial fixed points.

%%%%%%%%%%%%%%%%%%%%%%%%%%%%%%%%%%%%%%%%%%%%%%%%%%%%%%%%%%%%%%%%%%%%%%%%%%%%%%%%%%%%%%%%%%%%%%%%%%%%%%%%%%%%%%%%%%%%%%%%%%%%%%%%%%%%%%%%%%%%%%%%%%%%%%%%%%%%%%%%%%%%%%%%%%%%%%%%%%%%%%%%%%%%%%%%

\section{Conclusions}\label{conc}

In this paper we have shown that the structure of all known analytic solutions of open string field theory is inherited by tachyonic solutions of the effective spacetime action; compare eqs.~\Eq{rhoin} and \Eq{vsoin}. In both cases, the diffusion equation plays an important role for the construction of solutions. We have established a connection between two rather different setups (one exact and universal, the other approximate and spacetime dependent) which can be instrumental to finding other solutions in either framework. In particular, the construction of an exact supersymmetric solution with marginal deformations (in series and integral form) and of a non-perturbative non-marginal kink in the full theory will help to verify, and hopefully strengthen, the picture advanced here.

%%%%%%%%%%%%%%%%%%%%%%%%%%%%%%%%%%%%%%%%%%%%%%%%%%%%%%%%%%%%%%%%%%%%%%%%%%%%%%%%%%%%%%%%%%%%%%%%%%%%%%%%%%%%%%%%%%%%%%%%%%%%%%%%%%%%%%%%%%%%%%%%%%%%%%%%%%%%%%%%%%%%%%%%%%%%%%%%%%%%%%%%%%%%%%%%

%\bigskip
\begin{acknowledgments}
G.C. thanks S. Alexander for his kind hospitality at Haverford College, where this work was initiated. G.C. is partly supported by NSF grant PHY0854743, The George A. and Margaret M. Downsbrough Endowment and the Eberly research funds of Penn State. G.N.\ is partly supported by INFN of Italy.

\noindent {\bf Open Access.} This article is distributed under the terms of the Creative Commons
Attribution Noncommercial License which permits any noncommercial use, distribution,
and reproduction in any medium, provided the original author(s) and source are credited.
\end{acknowledgments}
 
%%%%%%%%%%%%%%%%%%%%%%%%%%%%%%%%%%%%%%%%%%%%%%%%%%%%%%%%%%%%%%%%%%%%%%%%%%%%%%%%%%%%%%%%%%%%%%%%%%%%%%%%%%%%%%%%%%%%%%%%%%%%%%%%%%%%%%%%%%%%%%%%%%%%%%%%%%%%%%%%%%%%%%%%%%%%%%%%%%%%%%%%%%%%%%%%

\end{document}